# Towards Refactoring of DMARF and GIPSY Case Studies – Team 8 SOEN6471-S14

## Guided by: Serguie A. Mokhov

Nitish Agrawal, Rachit Naidu, Sadhana Viswanathan, Vikram Wankhede, Zakaria Nasereldine, Zohaib S. Kiyani

Masters of Engineering in Software Engineering,Department of Computer Science And Engineering
Concordia University,Montreal,Quebec,Canada

I. Abstract

Of the factors that determines the quality of a software system is its design and architecture. Having a good and clear design and architecture allows the system to evolve (plan and add new features), be easier to comprehend, easier to develop, easier to maintain; and in conclusion increase the life time of the, and being more competitive in its market. In the following paper we study the architecture of two different systems: GIPSY and DMARF. This paper provides a general overview of these two systems. What are these two systems, purpose, architecture, and their design patterns? Classes with week architecture and design, and code smells were also identified and some refactoring's were suggested and implemented. Several tools were used throughout the paper for several purpose. LOGICSCOPE, JDeodoant, McCabe were used to identify classes with weak designs and code smells. Other tools and plugins were also used to identify class designs and relationships between classes such as ObjectAid (Eclipse plugin).

II. Introduction

Software application architecture is the process of defining a structured solution that meets all of the technical and operational requirements, while optimizing common quality attributes such as performance, security, and manageability. It involves a series of decisions based on a wide range of factors, and each of these decisions can have considerable impact on the quality, performance, maintainability, and overall success of the application[x]. The intent of this paper is to describe and discover the different design patterns and architectural designs that were used to develop two different systems GIPSY (General Intensional Programming System) and DMARF (Distributed Modular Audio Recognition Framework).

In the first section we introduce the two systems, DMARF and GIPSY, and we provide a general background about them and their uses. In the second section we discuss some requirements and design specifications (Personas, Actors, and Stakeholders, use cases and domain model). In the third section we show and discuss the actual system architecture and UML class diagrams of both systems. In the fourth section we try to discover the overall system design quality and identify the code smells found and suggest and show the implemented refactoring(s). In the fifth section we identify and design patterns that were used at the level of systems implementations (classes and their respective relationship in between each other.

## III. Background

### A. OSS Case Studies

#### DMARF

The Modular Audio Recognition Framework (MARF) is an open-source research platform and a collection of voice, sound, speech, text, and natural language processing (NLP) algorithms written in Java and arranged into a modular and extensible framework facilitating addition of new algorithms [10]. And whose application revolve around its recognition pipeline. Text Independent Speaker Identification is one of its application. The pipeline and application as they stand are purely sequential with little or no concurrency at all while processing the voice sample bulk [1]. The pipeline stages are the backbone of MARF through which all communication happens in chained manner. The pipeline consist of four different stages namely- sample loading, preprocessing, feature extraction, and training / classification. [5] The MARF pattern recognition pipeline is shown in figure 1 and showing the data flow and the transformation between different stages involved in MARF pipeline (Loaders, Preprocessing, Feature extraction, and Classification).

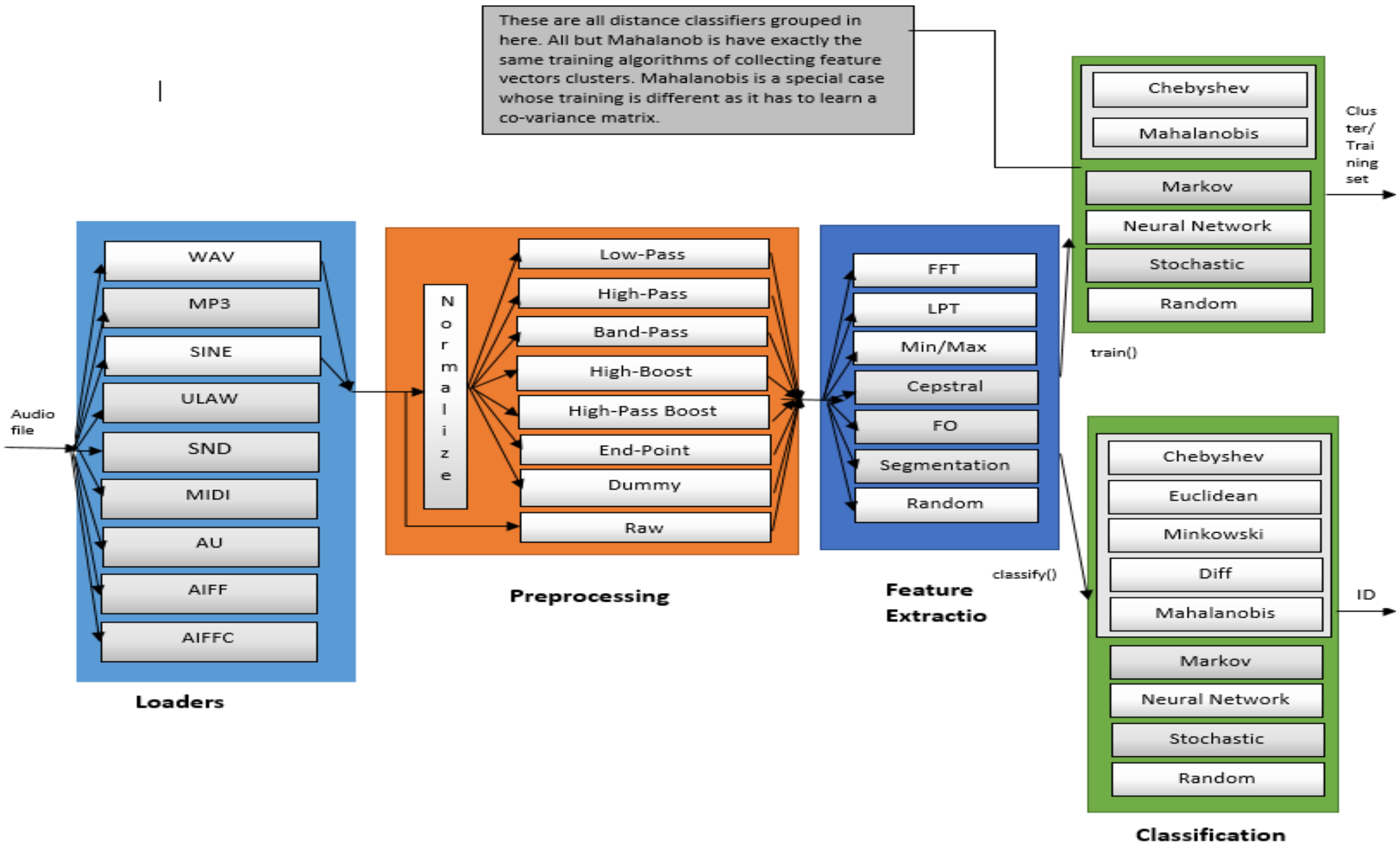

*Figure 1: Core MARF Pipeline Data Flow (Source: On Design and Implementation of Distributed Modular Audio Recognition Framework)*

DMARF, Distributed MARF, is based on the classical MARF whose pipeline stages were made into distributed nodes and later extended to be managed over SNMPv2 as shown below in figure 2. DMARF offers a number of service types [10]:

- Application Services
- General MARF Pipeline Services Sample Loading Services
- Feature Extraction Services

Training and Classification Services which are backed by the corresponding server implementations in CORBA, Java RMI, and Web Services XML-RPC.

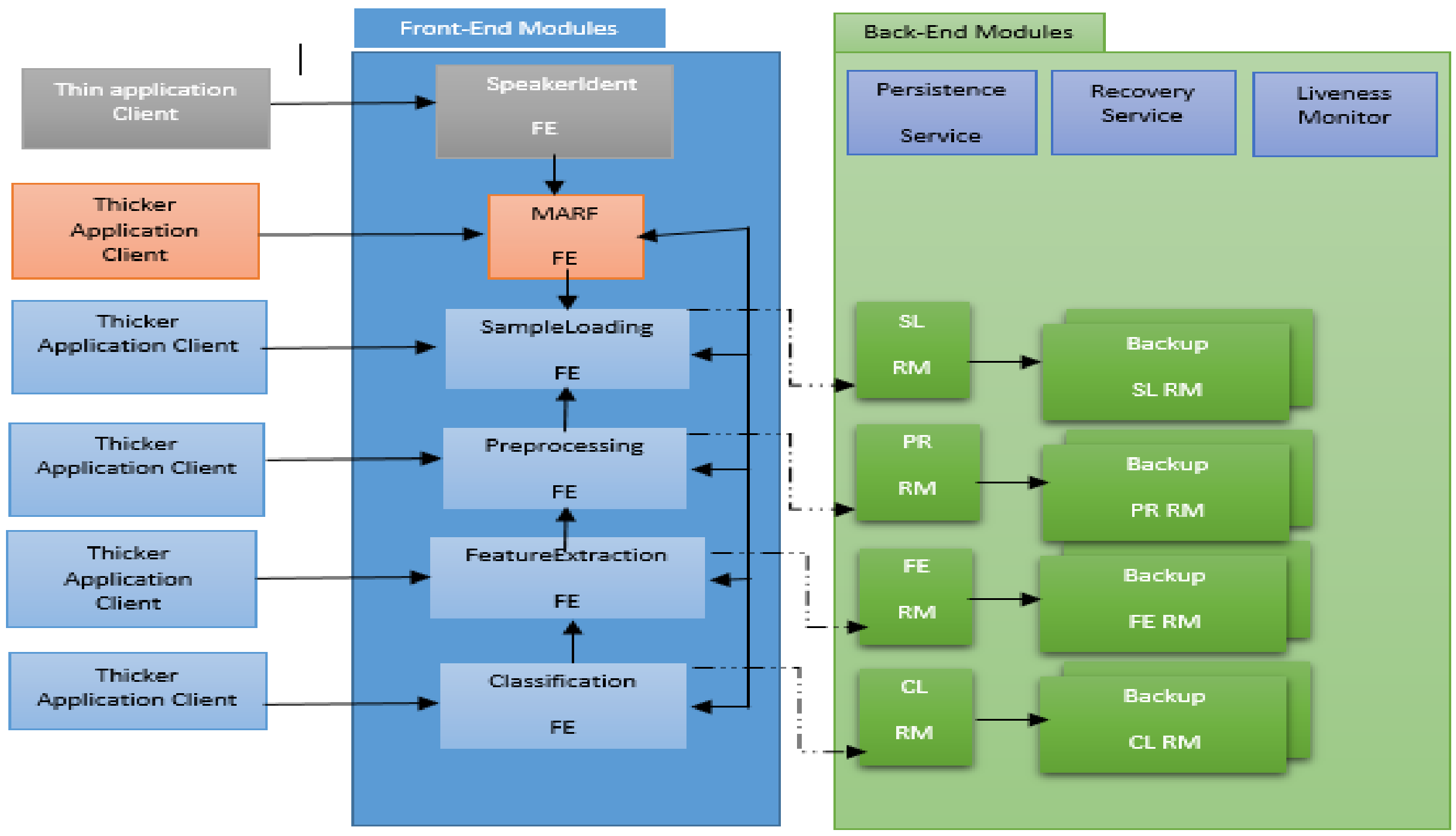

*Figure 2: The Distributed MARF Pipeline [23]*

DMARF is used to process audio, imagery and textual data by implementing pattern recognition, signal processing and natural language processing. The biometric applications use this framework. [3]

In Model View of system architecture DMARF application is divided into layers, Figure 2. Top level layer has a front end as well as a back-end however front end exists on client side and server side. It is either text-interactive or non-interactive client classes that connect and query the servers. In case of server side front end, the MARF pipeline is there along with the front end specific to the application and pipeline stage services. In execution view of system architecture there is hosting runtime environment of JVM in DMARF application while on server side the naming and implementation repository service must also be running. DNS and a web servlet container must be running for WS (Web Service) aspect application. Moreover Tomcat [4, 5] is used by DBS as a servlet container for MARF WS. Only JRE (Java Runtime Environment) is needed for RMI (Remote Method Invocation [2]) and COBRA (Common Object Request Broker Architecture) of client side. COBRA is one of the networking technologies used for remote invocation. RMI is the base line technology for remote method calls. All of the three (RMI, COBRA AND Web Services) main distributed technologies are used to implement MARF. These technologies can communicate through TCP or UDP.

Autonomic DMARF, A DMARF, is the self-managing capability in DMARF. The Autonomic System Specification Language (ASSL) is used to specify number of autonomic properties for DMARF such as self-healing, self-optimization, and self-protection. The ASSL framework implements these properties in the form of special wrapper Java code. [4] There are three major tier of ASSL i.e. the three major abstraction perspective:

- AS tier: Forms a global and general autonomic system perspective
- AS Interaction Protocol: Forms a communication protocol perspective.
- AE tier: Forms a unit level perspective

To achieve ADMARF there is a need to add autonomic computing behavior to the DMARF

behavior. Thus, the special autonomic (AM) manager is added to each stage of DMARF. Self-Healing is used to provide reliability by implementing replication technique to DMARF based system [7].

In case of local environment the self-protection of DMARF based system is not important but it crucial in case of global environments ran over the internet. [4] ASSL self-optimization model outlines the two major functional requirements. Data mirroring is used to optimize a lot computational effort. [6] There is automatic selection of the communication protocol due to use of self-optimization. The protocol is selected dynamically. [3] Thus, desired autnomicity in DMARF can be achieved by using ASSL.

Distributed Modular Audio Recognition Frame work is a fairly complex system composed of many levels of operational layers. The major problem is the use of DMARF is impossible in an unattended environment due to lack of design provisions that in turn necessitates the use of self-optimization feature.

GIPSY

The GIPSY (General Intentional Programming System) project is an ongoing research project developed at Concordia University. Its initial goal was to investigate on a general solution for the evaluation of programs written in the Lucid intentional programming family of languages using a distributed demand-driven evaluation model. In order to meet the flexibility goals of the project, the system has been designed using a framework approach integrating a Lucid compiler framework, as well as a demand-driven run-time system framework [27].

Using a framework approach, the GIPSY has been used to develop compilers for different variants of Lucid. Moreover, its flexible design also permits Lucid programs to use procedures defined in virtually any procedural language [11].

The GIPSY run-time system is a distributed multi-tier and demand-driven framework. It mainly consists of a set of loosely coupled software components enabling the evaluation of programs in a distributed demand-driven manner. The run-time system is composed of the following basic entities [28]:

A. A GIPSY tier is an abstract and generic entity. Each tier instance is a separate thread (one or more) that runs within a registered process, namely (GIPSY node). Tiers cooperate in a demand-driven mode of computation.
B. A GIPSY node is a registered process that hosts one or more GIPSY tier instances belonging to different GIPSY instance(s). Node registration is done through a manager tier called the GIPSY Manager Tier (GMT).
C. A GIPSY instance is a group of tier instances collaborating together to achieve program execution. A GIPSY instance can be executed across different GIPSY nodes (as shown in Figure 4). A GIPSY tier can be seen as a virtual network node and hosted on a GIPSY node. In such a network, the mapping between a GIPSY node and a physical node is made upon starting and registering the node through the GMT [28].

The GIPSY architecture inherits some of the peer-to-peer network architecture principles, such as:

- No single-point of failure: node failure does not mean system failure
- Nodes and tiers can join/leave the network by adding/removing them on the fly
- Demands are transmitted without knowing where they will be processed or stored

Available nodes and tiers can be affected at run-time to the execution of any GIPSY program while other nodes and tiers could be computing demands for different programs [28].

As it was mentioned above, the GIPSY has a multi-tier architecture where the execution of the GIPSY programs is divided into four different tasks assigned to separate tiers. The GIPSY processes communicate with each other through demands [13]. In In GIPSY, the notion of demand states for a request for the value of a program identifier in a specific context of evaluation. The GIPSY programs are evaluated by using a demand-driven lazy evaluation scheme [29]. The demands could be intensional, procedural demands, resource demands, and system demands.

## GIPSY ARCHITECTURE

GIPSY is a distributed system, designed as a modular collection of frameworks where components related to the development (RIPE, a

- **Program Compilation**

It is compiled in two stages, first the GIPSY program is translated into C, and then the The source code consists of two parts the Lucid part that defines the dependencies and the

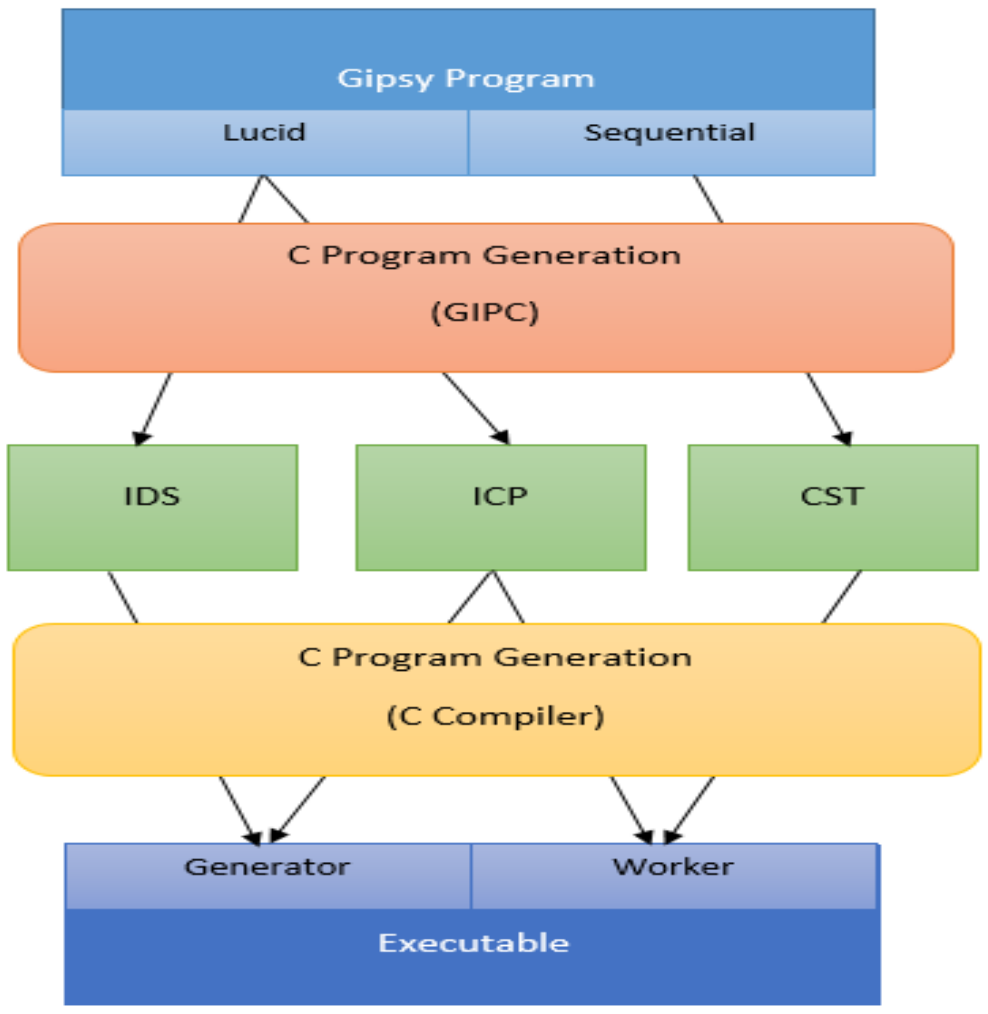

*Figure 3: Gipsy Program Compilation Process.png*

The Lucid part is compiled into Intentional data dependency units (IDU). Data communication procedures are also generated by GIPC yielding a set of intentional communication procedures (ICP). The sequential functions defined in the

Run-time Integrated Programming Environment), compilation (GIPC, the General Intensional Programming Compiler), and execution (GEE, the General Eduction Engine) of Lucid programs are decoupled to allow easy extension, addition, and replacement of the components. GIPSY has a collection of compilers under the GIPC framework and the corresponding run-time environment under the eduction execution engine (GEE) among other things that communicate through the GEE Resources (GEER). These two modules are the major primary components for compilation and execution of intensional programs [27].

**GIPC:**

resulting C program is compiled in a standardized manner.

sequential part that defines granular sequential computation units (See fig 6.) [16].

second part of GIPSY program are translated into C code using the second stage C compiler syntax yielding C Sequential Threads (CST)

- **GEE**

Gipsy basically uses a demand driven model of computation the process is called Eduction. Every demand generates a procedure call which is computed either locally or remotely. Every computed value is placed in a ware house and is used from there instead of computing a new one, thus reducing the overhead of time.

- **RIPE**

It basically shows the dataflow of the runtime programming environment corresponding to the Lucid part of the gipsy program. The GIPSY run-time system is a distributed multi-tier and demand-driven framework. It mainly consists of a set of loosely coupled software components enabling the evaluation of programs in a distributed demand-driven manner.

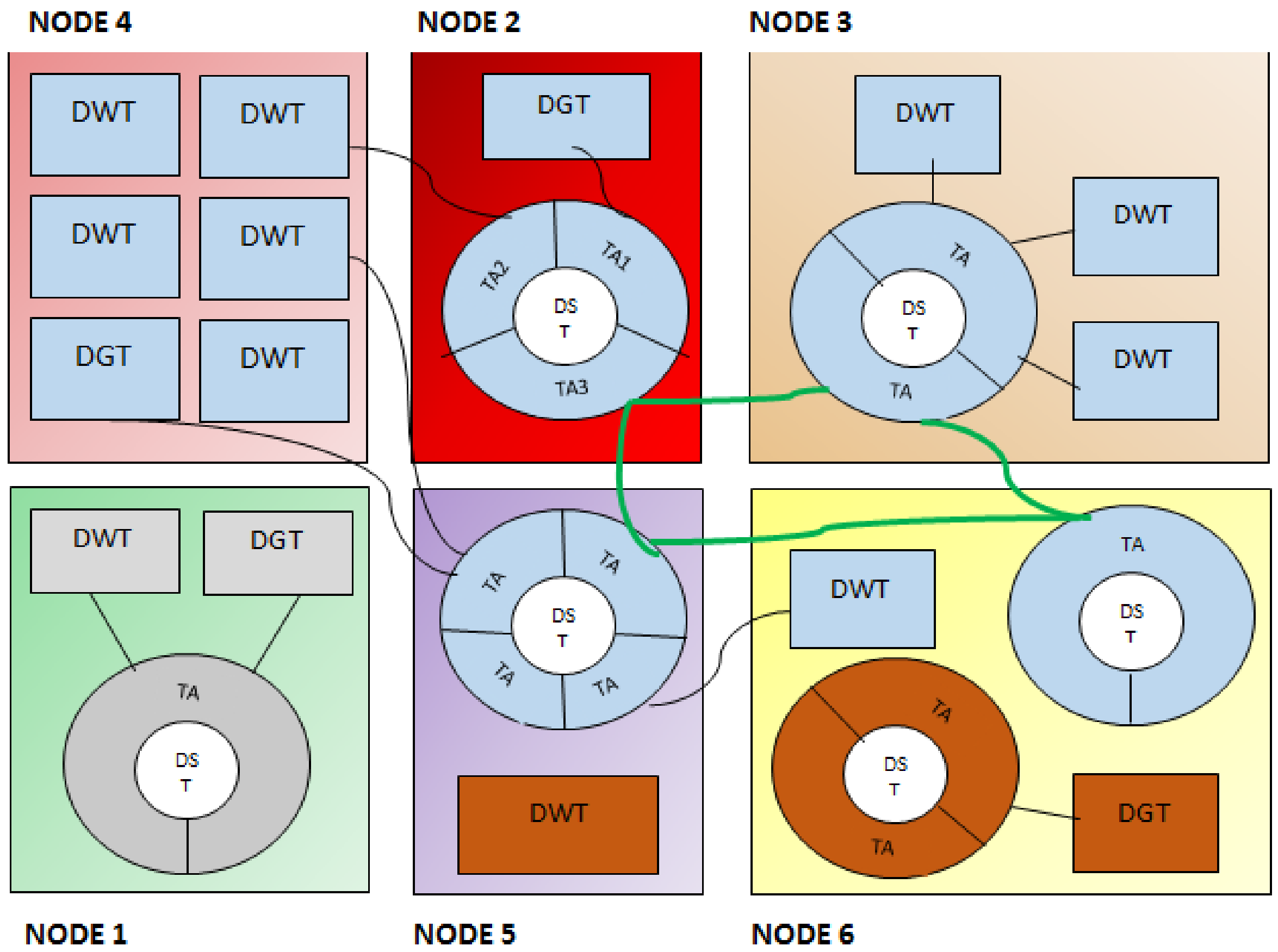

*Figure 4: GIPSY is a Multi-tier Architecture*

GIPSYs multi-tier architecture is composed of 4 main tiers [28]:

(a) A Demand Store Tier (DST) that acts as a middleware between tiers in order to migrate demands, provides persistent storage of demands and their resulting values, and exposes Transport Agents (TAs) used by other tiers to connect to the DST;

(b) A Demand Generator Tier (DGT) that generates demands according to the declarations and resources generated for the program being evaluated.

(c) A Demand Worker Tier (DWT) which processes demands by executing method defined in such a dictionary. The DWT connects to the DST, retrieves pending demands and returns back the computed demands to the DST;

(d) A General Manager Tier as its name implies (see Figure 4 above), locally and remotely controls and monitors other tiers (DGT, DWT and DST) by exchanging system demands. Also, the GMT can register new nodes, move tier instances from one node to another, or allocate/de-allocate tier instance from/on a registered node. [27] Of programs written in the Lucid intentional programming family of languages using a distributed demand-driven evaluation model. In order to meet the flexibility goals of the project, the system has been designed using a framework approach integrating a Lucid compiler framework, as well as a demand-driven run-time system framework. The GIPSY run-time system is a distributed multi-tier and demand-driven framework composed of 4 main tiers:

(a) A Demand Store Tier (DST) that acts as a middleware between tiers in order to migrate demands;

(b) A Demand Generator Tier (DGT) that generates demands;

(c) A Demand Worker Tier (DWT) which

processes demands;

(d) A General Manager Tier which locally and remotely controls and monitors other tiers (DGT, DWT and DST). Also, the GMT can register new nodes, move tier instances from one node to another, or allocate/de-allocate tier instance from/on a registered node. AGIPSY, Autonomic GIPSY, it is to GIPSY as ADMARF is to DMARF. It is the self-managing capability in GIPSY and its foundation is also a model built with the ASSL.

In this section we calculate the following for both of the source codes of DMARF and GIPSY:

- Number of Java files
- Number of Classes
- Number of Methods
- Number of lines of Java codes

We have mainly used the Eclipse InCode plugin to get the readings for the DMARF and we have used the Eclipse Code-Pro and InCode Plugins to get those of GIPSY as summarized in the below table:

| Attribute | Estimated values(DMARF) | Software Used | Estimated Values (Gipsy) | Software Used |
|---|---|---|---|---|
| Number of Java Files | 1024 | Eclipse (InCode Plugin) | 602 | Eclipse (CodePro and InCode Plugin) |
| Number of Classes | 216 | Eclipse(Incode Plugin) | 702 | Eclipse (CodePro and Incode Plugin) |
| Number of Methods | 2424 | Eclipse(InCode Plugin) | 6468 | Eclipse (CodePro and Incode Plugin) |

*Table 1: Initial Estimation Result*

## Summary

MARF, Modular Audio Recognition Framework, is an open-source research platform and a collection of voice, sound, speech, text, and natural language processing (NLP) algorithms written in Java and arranged into a modular and extensible framework facilitating addition of new algorithms and whose application revolve around its recognition pipeline. The pipeline stages are the backbone of MARF through which all communication happens in chained manner. The pipeline consist of four different stages: sample loading, preprocessing, feature extraction, and training / classification. DMARF, Distributed MARF, is based on the classical MARF whose pipeline stages were made into distributed nodes and later extended to be managed over SNMPv2. ADMARF is an extension of DMARF having the self-managing capability in DMARF. ADMARF foundation is a model build with ASSL. The ASSL consists of three main tiers: AS (Autonomic System) Tier, AS Interaction Protocol (ASIP), and AE (Autonomic Elements) Tier.

GIPSYs, General Intentional Programming System, initial goal was to investigate on a general solution for the evaluation.

IV. Requirements and Design Specifications

A. Personas, Actors, and Stakeholders

DMARF Actors

| Actor Name | Definition |
| --- | --- |
| Application | The system application on a whole |
| Classifier | The resultant from feature extractor is then classified which is composed of training and classify concepts |
| Feature extractor | Extracts important features of the preprocessed input |
| Preprocessor | Used to normalize data provided as input |
| Sample loader | To read audio information from a saved voice sample, a special sample-loading component had to be implemented in order to load a sample[24] |
| User | Any person using the system, can be an end user, researcher, system analysts or developer. |

*Table 2: DMARF Actors*

GIPSY Actors

| Actor Name | Definition |
| --- | --- |
| Compiler | Generates the result from the data provided by the evaluator |
| Evaluator | Evaluates the data sent by the processor, depending upon the demand. |
| Network | Provides peer-to-peer architecture to the GIPSY architecture |
| Node | A GIPSY node is a registered process that hosts one or more GIPSY tier instances belonging to different GIPSY instance(s). |
| Processor | Retrieves data from the nodes and tiers and sends to the evaluator |
| Tier | A GIPSY tier is an abstract and generic entity that cooperate in a demand-driven mode of computation |
| User | Any person using the system, can be an end user, researcher, system analysts or developer. |

*Table 3: GIPSY Actors*

Stakeholders

| Stakeholders Name | Definition |
| --- | --- |
| Concordia University | Exploring research both on GIPSY and DMARF |
| System Admin | Administrator of the system having all the system authorization |
| User | Any person using the system, can be an end user, researcher, system analysts or developer. |

*Table 4: DMARF and GIPSY Stakeholders*

DMARF Persona

<table>
<tr>
<td>Chris Moran<br>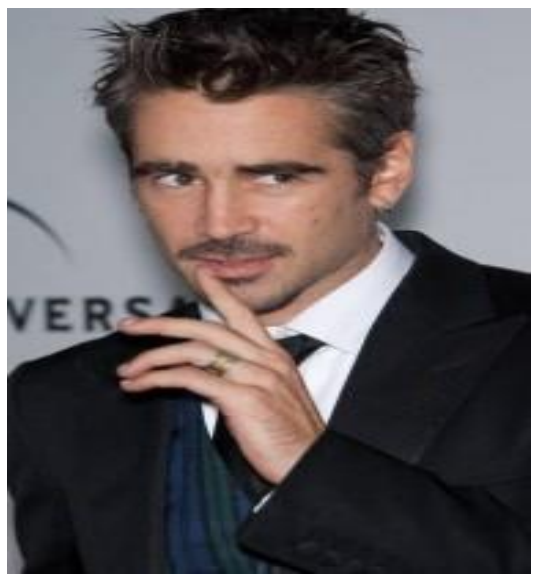</td>
<td><u>Background</u><br>Age: 40years<br>Occupation: Research Analyst<br>Company: Hybris (Canada)<br>Technology Level: High.</td>
</tr>
<tr>
<td><u>Main Points</u><ul><li>He is a research analyst at Hybris, Canada.</li><li>He works as a freelance mobile application developer.</li></ul></td>
<td rowspan="4"><u>Description</u><br>Chris is from Ireland working as research analyst at Hybris, Canada. He completed his Masters of Software engineering from Concordia university.<br><br>The main objective is to distribute the stages as services as well as stages that are not directly present.<br><br>Extend the original DMARF with the Web Services (WS) implementation such that its architecture and semantics are compatible to that of the already fully implemented RMI and CORBA services.[25]<br><br>Whenever Chris tries to process sample voice in bulk, there is lack of concurrency in the output. Implementation not flexible enough to co-exist for interoperability and platform-independence of Java and CORBA. Due to lack of design provisions Use of DMARF is impossible in an unattended environment.<br><br>The Autonomic System Specification Language (ASSL) is used to specify number of autonomic properties for DMARF. The case study focuses on three such a properties namely self-healing, self-optimization, and self-protection. The ASSL framework implements these properties in the form of special wrapper Java code. [26]</td>
</tr>
<tr>
<td><u>Goal</u><ul><li>Make the pipeline distributed and run on a cluster.</li><li>Extend the original DMARF with the Web Services (WS) implementation.</li></ul></td>
</tr>
<tr>
<td><u>Frustration & Pain Points</u><ul><li>There is little or no concurrency during voice sample bulk processing.</li><li>Implementation not flexible enough</li><li>Use of DMARF is impossible in an unattended environment due to lack of design provisions.</li></ul></td>
</tr>
<tr>
<td><u>Scenarios</u><ul><li>To process audio, imagery and textual data by implementing pattern recognition, signal processing and natural language processing.</li><li>Self- healing, self-optimization, self-protection.</li></ul></td>
</tr>
</table>

*Table 5: DMARF Persona*

GIPSY Persona

| | |
|---|---|
| **Mark Anthony**<br>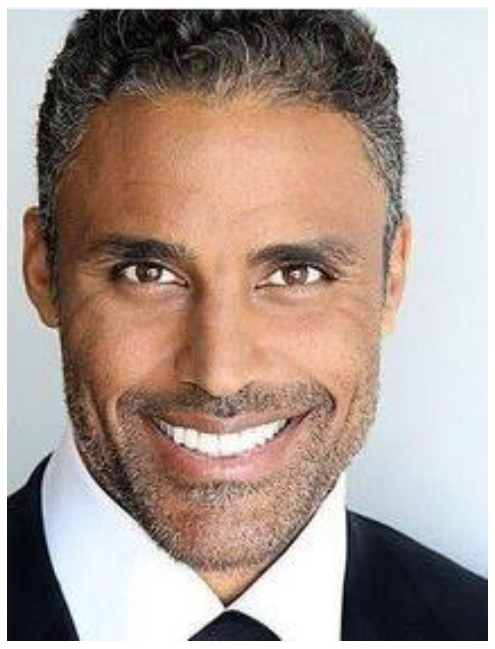 | **Background**<br>**Age:45 years**<br>**Occupation:** Senior Software Designer<br>**Company: BioCorps (Canada)**<br>**Technology Level:** High |
| **Main Points**<br>• He works as a senior software designer at the BioCorps.<br>• Researching about GIPSY to find out methods of better translation of textual inputs to graphical outputs and vice versa. | **Description**<br>Mark is 45 year old lives in Montreal he have been working at BioCorps for 10 years. He is always interested in researching about various methods and technologies.<br>He have been working on GIPSY to find out methods of better translation of graphical to textual outputs or vice versa.<br>For the system to be easily understandable, we should provide an intelligent interface. For instance, an application in the scientific domain, the system should be efficient and stable and fault resistant.<br>Mark is suggesting potential solution as usage of dataflow graph which will translate the textual version of GIPL program to make it more understandable. The translator has to be flexible so that system can deal with any translation between a new SIPL and GIPL. |
| **Goal**<br>• should provide a translator that can translate textural version into graphic version<br>• system provides a translator which can translate graphic version into textual version | |
| **Frustration & Pain Points**<br>• Textual version of the GIPL programs is considered very hard to decipher.<br>• provide intelligent interface | |

| **Scenarios** | |
|---|---|
| • Dataflow graph notation makes it much easier to understand textual version of GIPL program.<br>• Translator has to be flexible. | |

*Table 6: GIPSY Persona*

B. Use Cases

DMARF

| **Use Case DMARF** | |
|---|---|
| Scope | System under design |
| Level | User Level |
| Primary Actor | Any User |
| Secondary Actor | 1. Autonomic researcher<br>2. Professor |
| Stakeholders and Interest | • Analyst: A person who intend to use the system to analyze audio using SpeakerIdentApp of DMARF.<br>• Professor.<br>• Organizations: Who are funding the project? |
| Preconditions | Audio or voice sample(s) provided to the server, making a DMARF-implementing network is valid |
| Post condition | Results generated based upon the inputs, identifying the speaker, their gender, accent, spoken language etc. |
| Main success scenario | 1. User indicates the system to have a biometric subject identification and analysis.<br>2. System allows user to upload his/her file to analyze on speakerIdentApp.<br>3. User uploads his/her audio file.<br>4. System indicates that the file is successfully uploaded on speakerIdentApp.<br>5. SpeakerIdentApp forwards the file to MARF to recognize it.<br>6. MARF start recognitionPipeline and from SampleLoader, it loads some sample files<br>7. MARF sends the sample to preprocessing layer for concrete preprocessing.<br>8. Preprocessing layer normalizes the sample array and send it back to sample layer.<br>9. MARF take data from preprocessing and send it to feature extraction layer.<br>10. Feature extraction layer generates feature vector.<br>11. MARF classifies the features extraction with the help of classification layer.<br>12. Classification layer generates results<br>13. MARF gets results from the result layer<br>14. SpeakerIdentApp gets result from MARF |
| Extension/Alternatives scenario | **2(b) Uploaded format is incorrect:**<br>If the uploaded file format is invalid or the file is corrupted, the system should indicate that the file is invalid/corrupted.<br>**4(b) Uploaded file is not successful:**<br>If the upload of the file was not successful the system should prompt the user to re-upload the file again (redo step 3) |
| Special requirements | - User machine is connected to the server (DMARF database).<br>- Audio samples should be valid and of specific format (wav, mp3). |
| Technology and data Variations list | Cellphone, voice recorders, Audio files (mp3, wav etc.) CORBA,RMI |
| Open issue | N/A |

*Table 7: DMARF Use Case*

GIPSY

| **Use Case GIPSY** | |
|---|---|
| Scope | System under design |
| Level | User Level |
| Primary Actor | Developer |
| Secondary Actor | Network |
| Stakeholders and Interest | Developer: The developer wants be able to have the functionality of translating graph versions to textual version |
| Preconditions | Datagrah is valid |
| Post condition | 1. The system translates the graphic version (datagraph) into textual version (code).<br>2. The system compiles the generated code. |
| Main success scenario | 1. Developer draws the graphic version through the RIPE.<br>2. Developer supplies the system with the drawn graph.<br>3. GIPC reads the graph checking for syntax error in the graph drawn.<br>4. GIPC generates the textual version (code) of the provided graph (translated in Java).<br>5. GIPC compiles the program in the standard way.<br>6. GIPC generates the GEER (GEE Resource which is the stored compiled GIPSY program) which is a data dictionary storing all program identifiers, encapsulated with all ASTs generated at compile time.<br>7. GEER is fed to the Demand Generator (DG) by the GIPC<br>8. The DG makes a request to the warehouse (Data Storage DS) to see if this demand has already been computed.<br>9. The Data worker (DW) connects to the DST, retrieves pending demands and returns back the computed demands to the DS.<br>10. System returns back the result of the executed program to the user. |
| Extension/Alternatives scenario | **3(a) Supplied graph is invalid:**<br>If the supplied graph is invalid the system should generate a message stating that compilation error has occurred, please review the supplied graph. |
| Special requirements | LUCID compiler framework should be installed |
| Technology and data Variations list | Java (JDK, JVM, etc…), RMI, and CORBA should be available/installed |
| Open issue | N/A |

*Table 8: GIPSY Use Case*

C. Domain Model UML Diagrams
DMARF

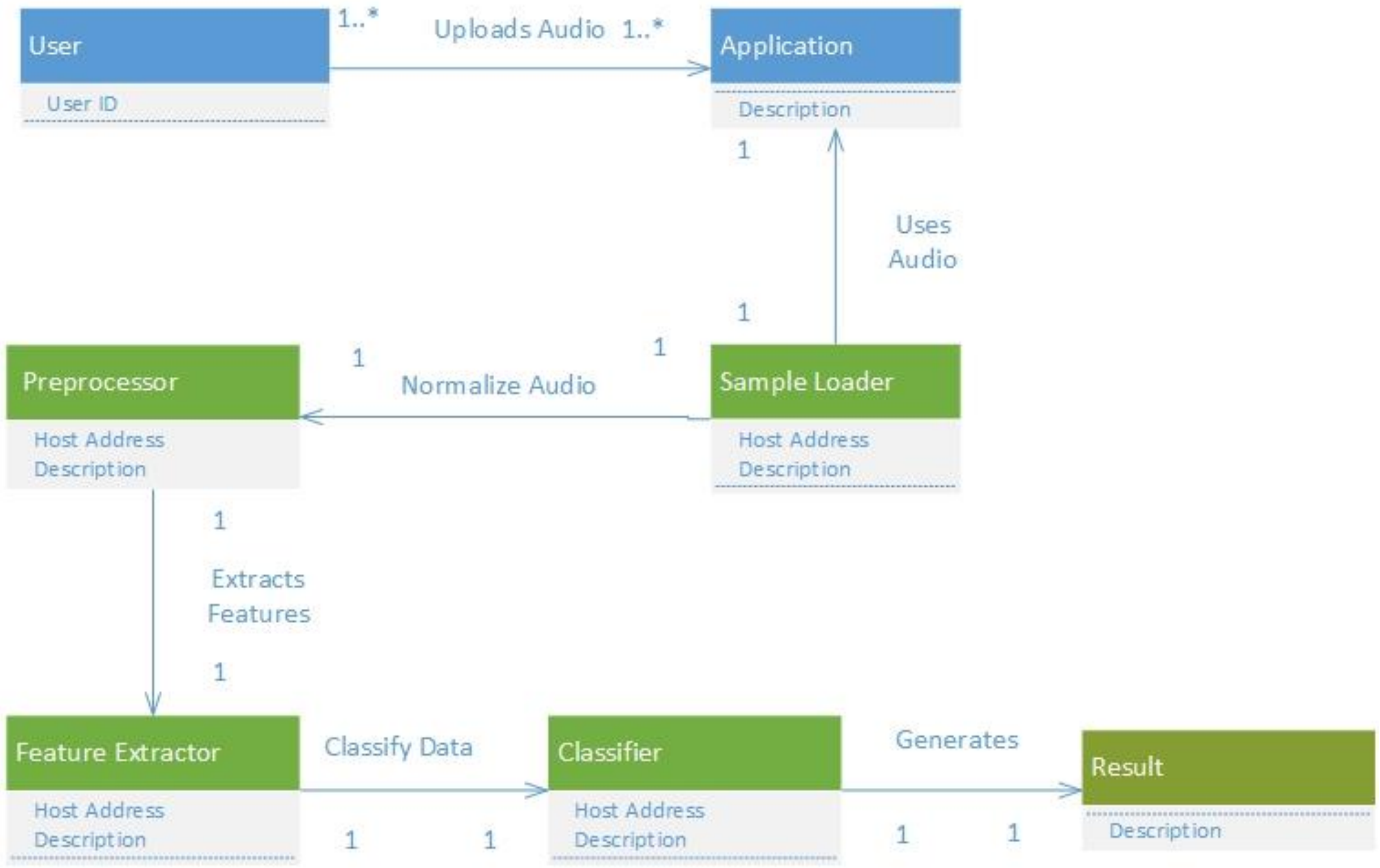

*Figure 4: DMARF Domain Model UML Diagram*

**Description:**

In this domain model we tried to cover all important concepts in an abstract manner. The user is a specified form of a person. The main attribute of user concept is User ID. User uploads the data using application service. Sample loader loads the data from application. It is composed of different types of loaders which is called at runtime.

Preprocessor is used to normalize data. The normalization process is done using various filters. The processed data is further passed to feature extractor which provides the important features. Finally, the data is classified using classification which is composed of training and classify concepts. The result is the output of this whole process.

In DMARF sample loader, preprocessor, feature extractor and classifier are situated on different hosts thus they consist of host address as a one of their attribute.

GIPSY

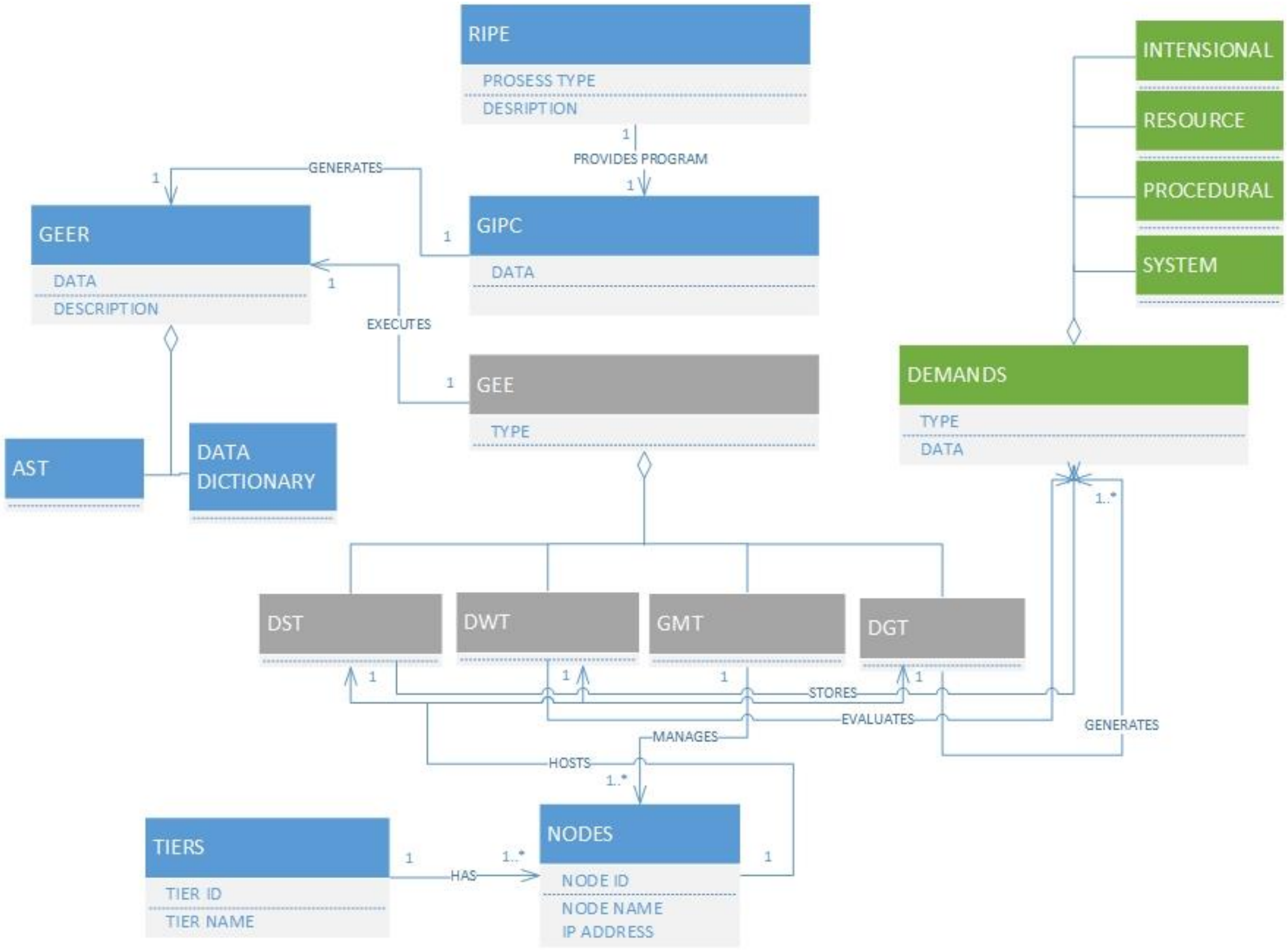

*Figure 5: GIPSY Domain Model UML Diagram*

**Description:**

In this domain model we have described high level implementation of gipsy so that it is understood by a normal user. GIPSY is a system composed of three main subsystems: RIPE, GIPC, and GEE. Usually RIPE provides GIPC with the program to be compiled. GIPC generates the GEER that is needed to be executed by the GEE. GEER is a composite of 2 components: AST (Abstract syntax tree) and the Data Dictionary. GEE is composed of 4 components/concepts: DST, DGT, GMT, and DWT. DGT generates demands and DWT evaluate demands and DST stores executed demands. Demands could be of four different types: Intentional, Resource, Procedural, and System. Also Gipsy consists of nodes and tiers. GIPSY tier is an abstract and generic entity. Each tier instance is a separate thread (one or more) that runs within a registered process, namely (GIPSY node). A GIPSY node on the other hand is a registered process that hosts one or more GIPSY tier instances belonging to different GIPSY instance(s). Node registration is done through a manager tier called the GIPSY Manager Tier (GMT). Moreover, nodes hosts the GEE components (DST, DGT, and DWT).

Fused DMARF-Over-Run-time Architecture (DoGTRA)

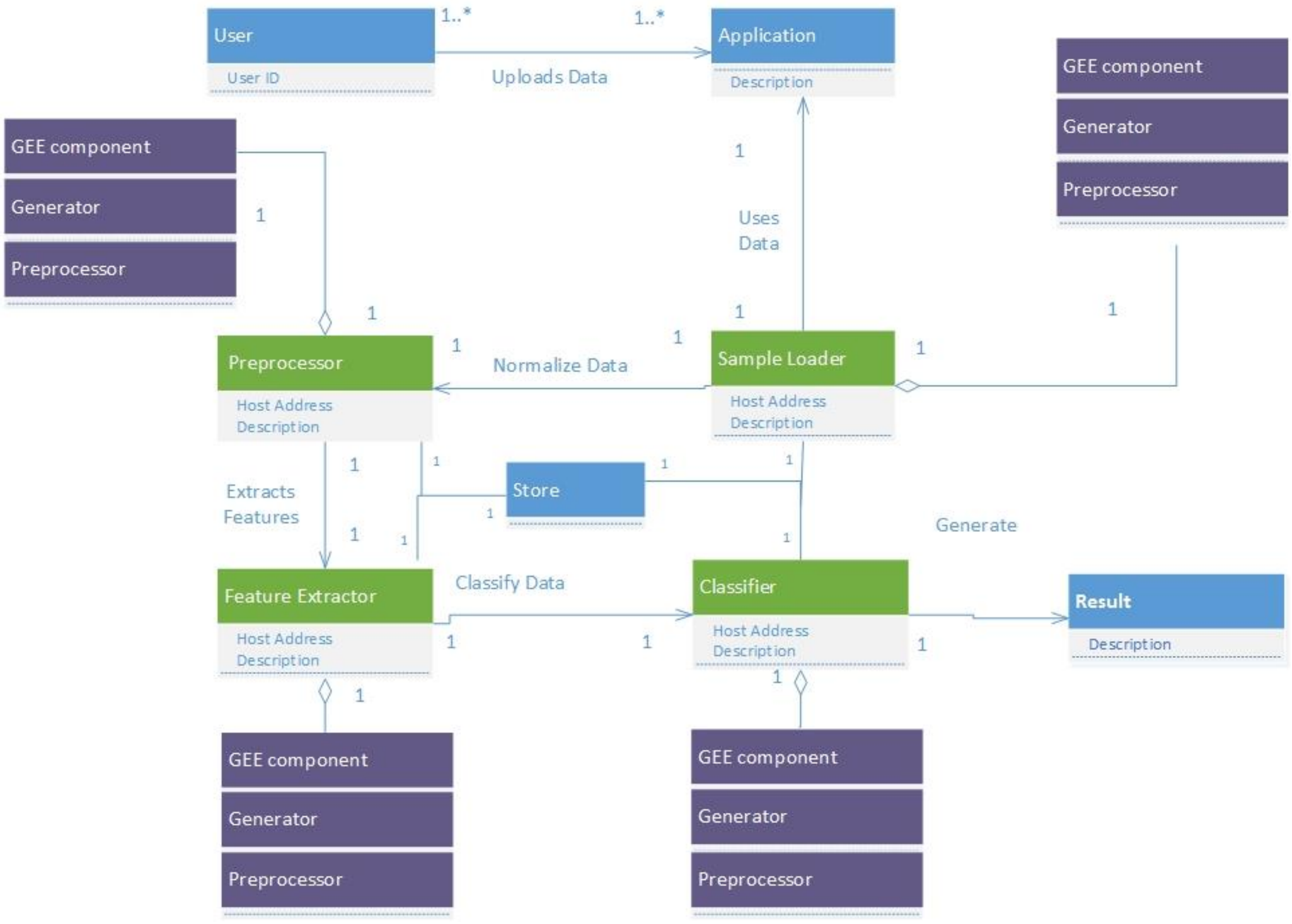

*Figure 6: Fused DMARF-Over-Run-time Architecture (DoGTRA)*

**<u>Description:</u>**

In this merged domain model we tried to combine advantages of DMARF and GIPSY. The atomicity feature is provided to DMARF at runtime using GEE (General Eduction Engine) multi-tier architecture for distributed computed instead of self-realization.

The DMARF pipeline stages will be divided into two categories: Demand generation, and demand execution. The first category, demand generation, will be done by the loaders and preprocessing. These two components will use and run over the demand generation tier of GIPSY (DGT). The second category, related to execution will contain Feature extraction and the classifiers. These two components usually process data and classify it. These components will be using and running over the GIPSY demand worker tier (DWT). Both categories will be communicating with the GIPSY storage tire (DST) for storing demands and the results of these demands. GIPSY's GMT will be used to manage the usage of the GIPSY components by the DMARF pipeline stages.

Therefore, the GEE multi-tier architecture of GIPSY is used by all pipelined stages of DMARF.

## D. Actual Architecture UML Diagrams

### DMARF

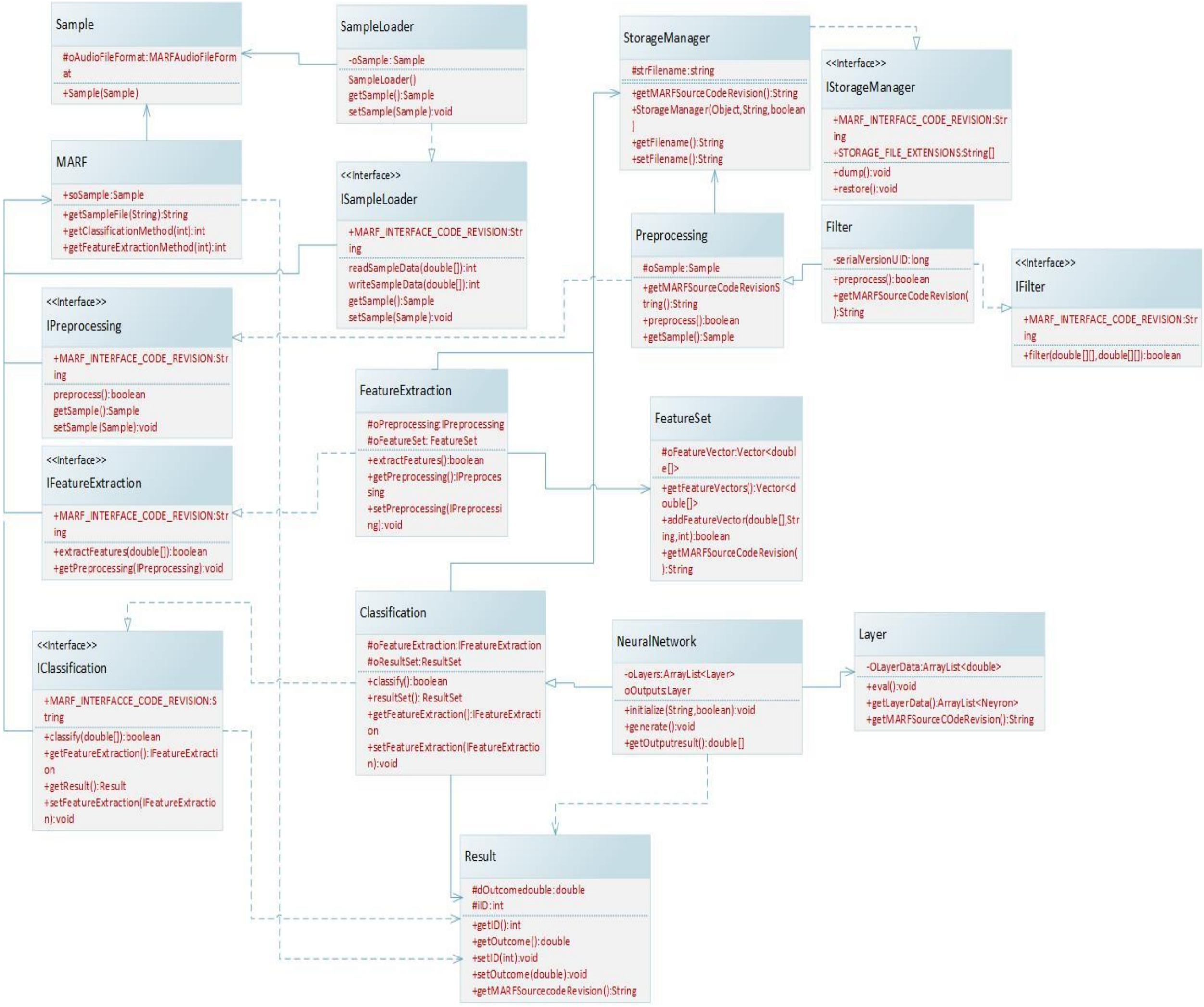

*Figure 7: DMARF UML Class Diagram*

**DMARF Class Diagram Description:**

The above diagram shows the list of main classes used in the class diagram of the DMARF use case that was mentioned previously. Moreover this class diagram describes the relationships between the classes and components of the DMARF system. In the above diagram we have four types of relationships between classes. The first one is directed association relationship, which is the most common type of relationships between classes. This relationship exists between the Storage manager and classification classes, between the interface classes (IClassification, IFeatureExtraction, IPreprocessing, and ISampleLoader) and the class MARF.

The second type is the implementation relationship. This type of relationships is found between classes and interfaces that implement these interfaces. In the above diagram this relationship is obvious from the naming of classes. You can find this relationship between the classes Classification and ICLassification, IFeatureExtraction and FeatureExtraction, IPreprocessing and Preprocessing, ISampleLoader and SampleLoader. In all these cases the extending class is implementing a method from the interface/extended class. For example, the Classification class is implementing the method classify() which is found in the interface IClassification.

The third type of relationships between classes in the diagram is the dependency. Dependency here is in the sense that the changes to one model element (class) will impact another model element. In the above diagram we have dependency from several classes/interfaces to the Result class. IClassification, MARF, and NeuralNetwork have dependency over Result. For example, the class Result stores intermediate results and the IClassification extracts it and sends to the Classification class.

The fourth type is the aggregation relationship. This relationship appears between Classification and NeuralNetworks, and Preprocessing and Filter. In these two cases the part class appears to be a part of the functionality of the aggregate class in domain model but here it is separated into another class.

Generally speaking in DMARF the main conceptual classes from the domain model are also found in the design class diagram. The Preprocessor, Feature Extractor, Classifier, Result, Sample Loader are all found in the class diagram. The Preprocessor maps to the class Preprocessor, Feature Extractor maps to FeatureExtraction, Classifier maps to Classification, Result maps to Result, Sample Loader maps to SampleLoader.

In the class diagram, the solution domain, we can found that more classes are found. Usually while system implementation (design and coding) new classes and concepts appear for several technical reasons. In DMARF case we can see that interfaces (IClassification, IFeatureExtraction, Preprocessing, and ISampleLoader) are added and these are implementation specific entities that don’t exist in real world or problem domain world. Also some concepts in domain model are split into more than one entity/class in the solution domain for design purpose (class is very huge with big number of functionalities, low cohesion between class components, etc…). This situation for example can be found in the case of Classification class where other components such as Layer and NeuroNetwork are added. And in another case we have Preprocessing and Filter where in real world filtering is part of preprocessing but here it is separated also for design reasons.

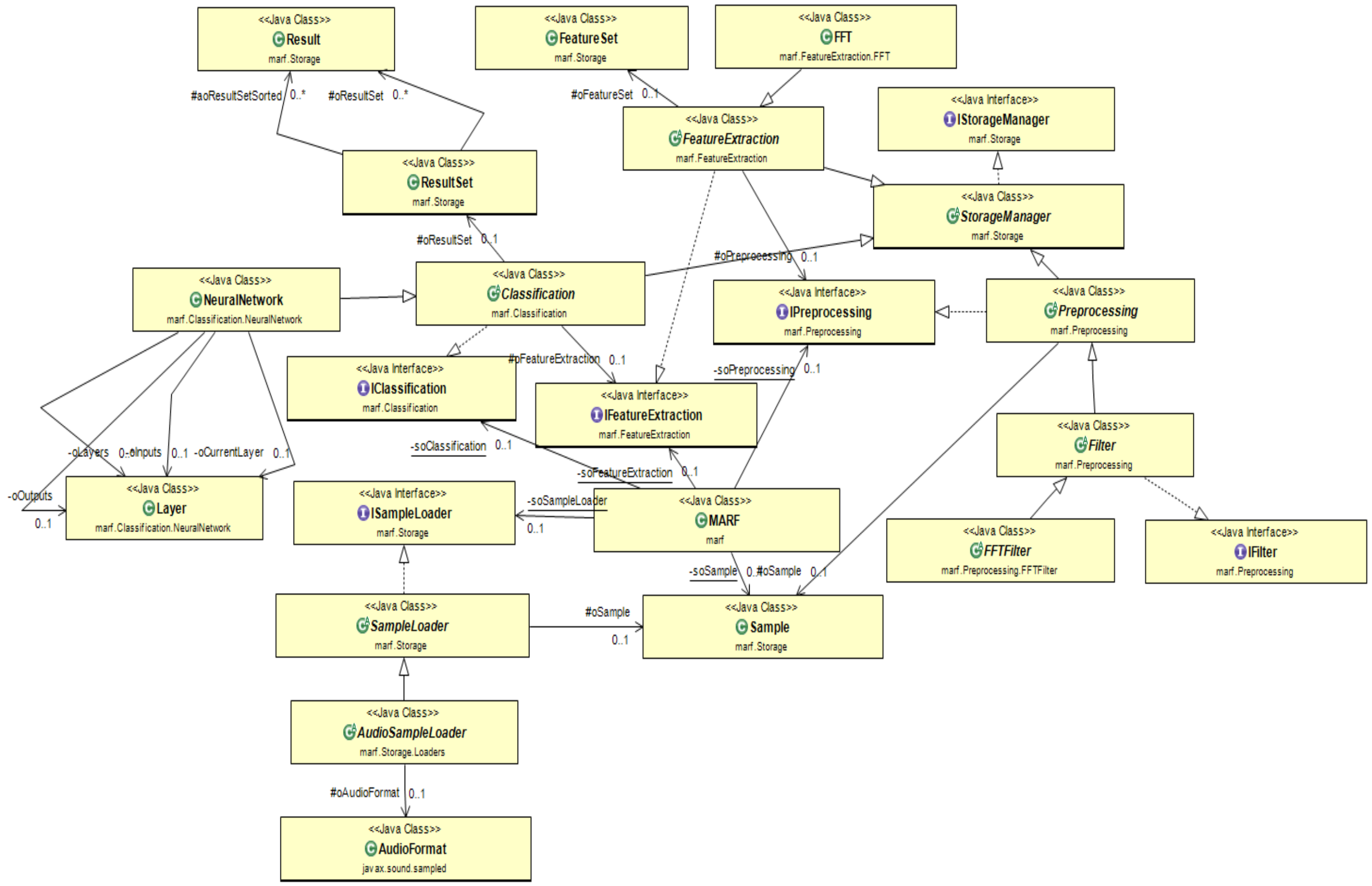

*Figure 8: DMARF Reverse Engineered Class Diagram*

GIPSY

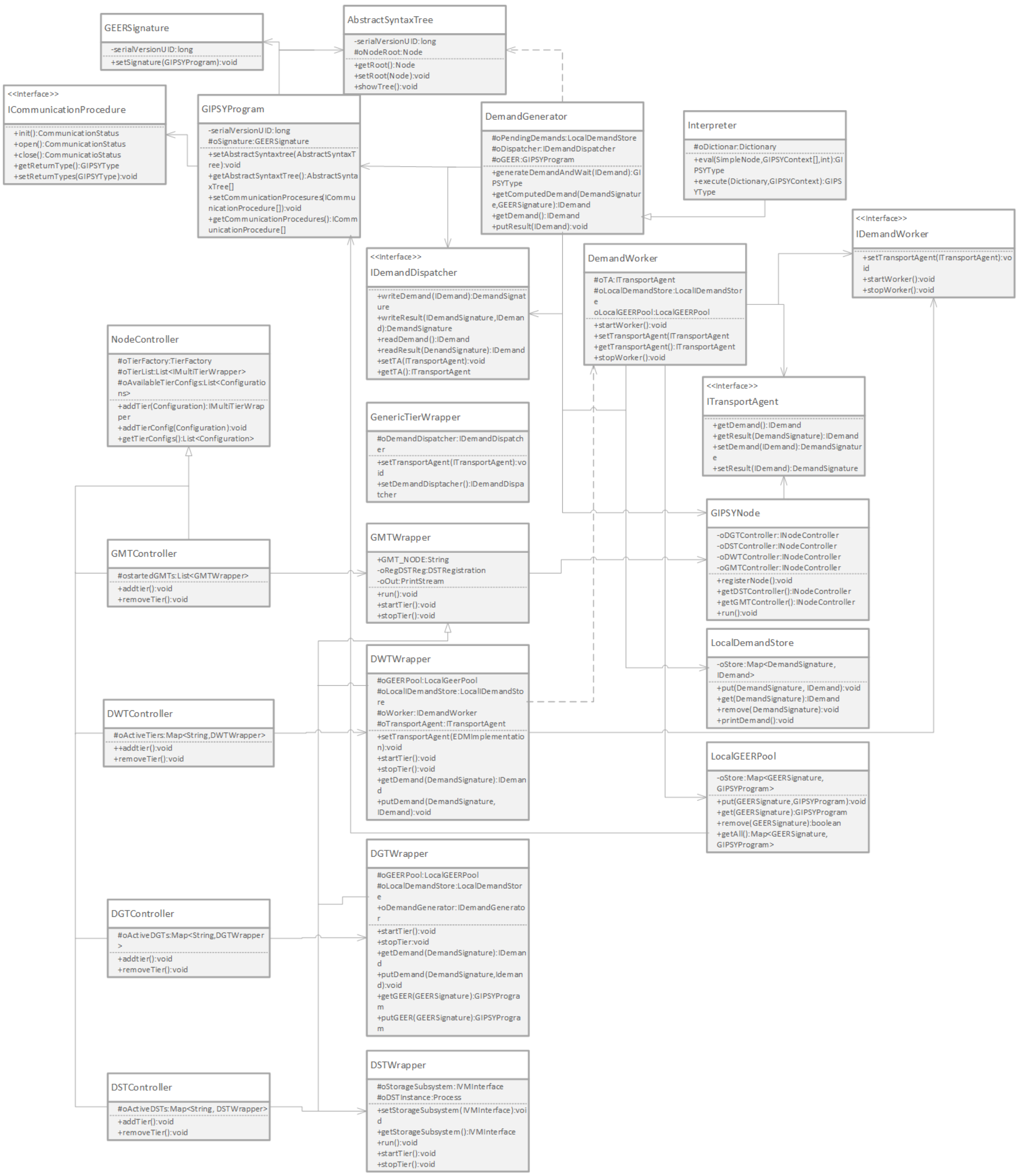

*Figure 9: GIPSY UML Class Diagram*

## GIPSY Class diagram Description:

Above is the class diagram that shows the list of classes of interest when it comes to the GIPSY use case mentioned previously in the above sections. Moreover this class diagram describes, as the DMARF's class diagram, the relationships between the classes and components of the GIPSY system. In the GIPSY class diagram we have five types of relationships between classes. The first one is directed association relationship. This relationship exists heavily between classes and of these we mention the following:

1. GIPSYProgram: with GEERSignature and AbstractSyntaxTree and the.
2. DemandGenerator: with GIPSYProgram, GIPSYNode, and LocalDemandStore.
3. Between Controller classes and Wrapper classes (Such as DSTController and DSTWrapper).

The second type is the implementation relationship. This type of relationships is also found between classes and interfaces that implement these interfaces. In the above diagram this relationship is obvious from the naming of classes. You can find this relationship between the classes DemandWorker and IDemandWorker. Other implementation relationships are also found where classes are not related to each other in terms of naming (no similarity in naming) such as ICommunicationProcedure and GIPSYProgram.

The third type of relationships between classes in the diagram is the dependency. In the above diagram we have dependency between several classe. For example we have dependency between DemandGenerator and the

AbstractSyntaxTree. Where the DemandGenerator depends on the generated AbstractSyntaxTree to start generation of demands

The fourth type is the aggregation relationship. This relationship appears between GEE and the GMT, DST, DWT, and DGT. These four components together form the GEE subsystem.

The Fifth type is inheritance between some super-classes and sub-classes. This relationship is shown between NodeController and GMTController, DWTController, DGTController, and DSTController where the later four components inherit most of their behavior from the super-class NodeController.

As in DMARF, the same here applies in GIPSY. The main conceptual classes from the domain model are also found in the design class diagram (you can see that in the above Figure 9 and Figure 11 below). The GEE, RIPE, GIPC, GMT, DST, DGT, GEER, AST are also found in the class diagram. Also these concepts map to classes with the same naming in the class diagram (AST maps to AbstractSyntaxTree, GEE maps to GEE, GEER maps to GEER and so on so forth).

In the class diagram, the solution domain, we can found that more classes are found (same case as DMARF). For the same technical details stated before new classes and concepts appear for several in this domain without existing in real world situation. Similarly interfaces are introduced as objects that does not have existence in real world. Also some concepts in domain model are split into more than one entity/class in the solution domain.

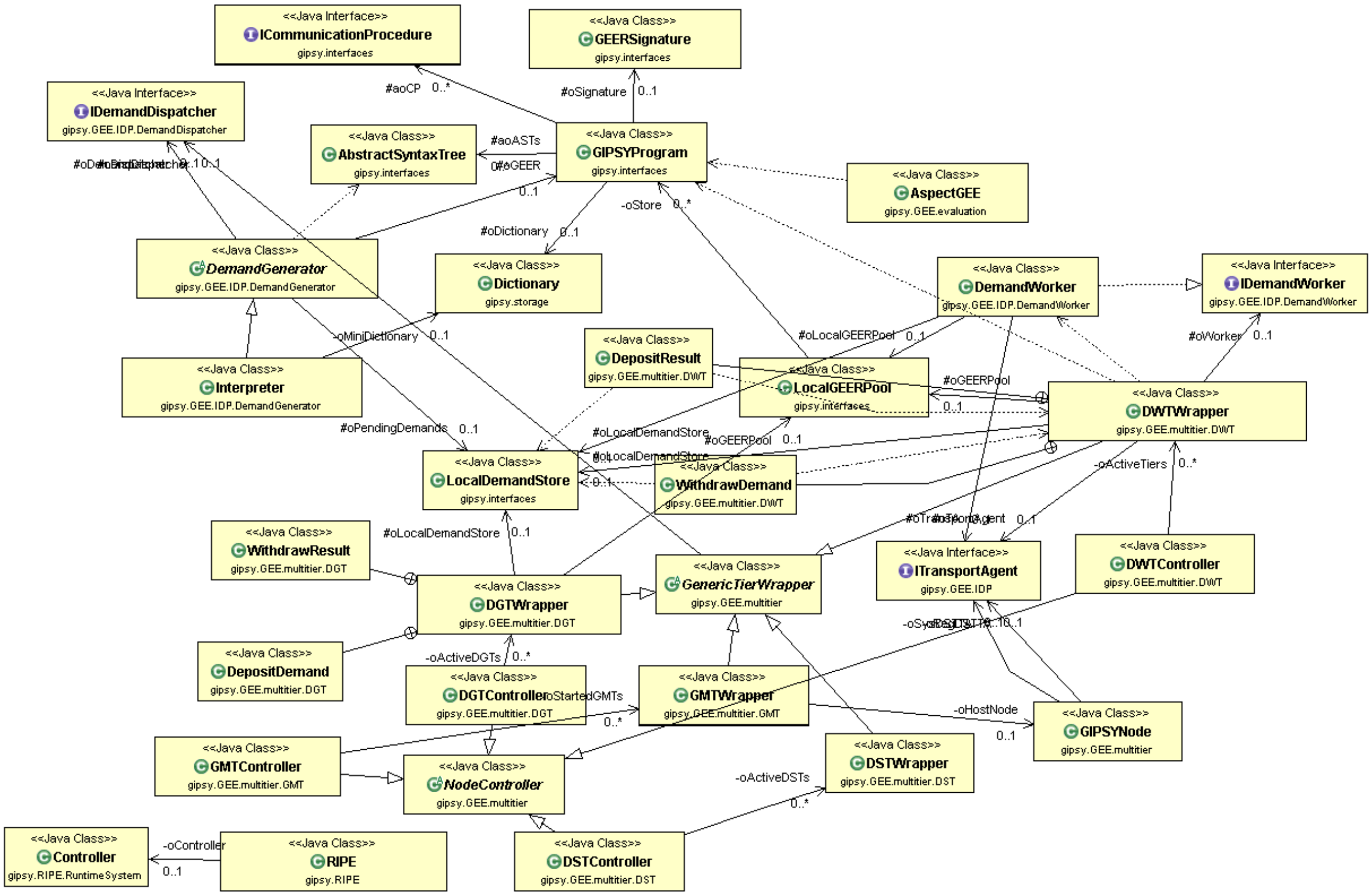

*Figure 10: GIPSY Reverse Engineered Class Diagram*

# V. Methodology

## A. Refactoring

### 1. Identification of Code Smells and System Level Refactoring DMARF & GIPSY

| **List of poor Classes in DMARF** |
|---|
| marf.Classification.NeuralNetwork. NeuralNetwork |
| marf.Classification.Stochastic.ZipfLaw |
| marf.Configuration |
| marf.MARF |
| marf.math.ComplexMatrix |
| marf.math.Matrix |
| marf.nlp.Parsing.GrammarCompiler.Grammar |
| marf.nlp.Parsing.GrammarCompiler.GrammarCompiler |
| marf.nlp.Storage.Corpus |
| marf.Stats.StatisticalEstimators.StatisticalEstimators |
| marf.Storage.ResultSet |
| marf.Storage.StorageManager |
| marf.util.Arrays |

*Figure 11: List of poor Classes in DMARF*

| **List of POOR Classes in GIPSY** |
|---|
| gipsy.Configuration |
| gipsy.GEE.IDP.DemandGenerator.jini.rmi.JiniDemandDispatcher |
| gipsy,GEE.IDP.DemandGenerator.jini.rmi.JINITA |
| gipsy.GEE.IDP.DemandGenerator.jms.JMSTransportAgent |
| gipsy.GEE.IDP.demands.Demand |
| gipsy.GEE.multitier,GIPSYNode |
| gipsy.GEE.multitier.GMT.GMTWrapper |
| gipsy.GIPC.DFG.DFGAnalyzer.DFGParser |
| gipsy.GIPC.DFG.DFGAnalyzer.DFGParserTokenManager |
| gipsy.GIPC.DFG.DFGGenerator.DFGCodeGenerator |
| gipsy.GIPC.DFG.DFGGenerator.DFGTranCodeGenerator |
| gipsy.GIPC.GIPC |
| gipsy.GIPC.intensional.Generic Translator.TranslationParser |
| gipsy.GIPC.intensional.GIPL.GIPLParser |
| gipsy.GIPC.intensional.GIPL.GIPLParserTokenManager |
| gipsy.GIPC.intensional.SIPL.ForensicLucid.ForensicLucidParser |
| gipsy.GIPC.intensional.SIPL.ForensicLucid.ForensicLucidParserTokenManager |
| gipsy.GIPC.intensional.SIPL.ForensicLucid.ForensicLucidSemanticAnalyzer |
| gipsy.GIPC.intensional.SIPL.IndexicalLucid.IndexicalLucidParser |
| gipsy.GIPC.intensional.SIPL.IndexicalLucid.IndexicalLucidParserTokenManager |
| gipsy.GIPC.intensional.SIPL.JLucid.JGIPLParser |
| gipsy.GIPC.intensional.SIPL.JLucid.JGIPLParserTokenManager |
| gipsy.GIPC.intensional.SIPL.JLucid.JIndexicalLucidParser |
| gipsy.GIPC.intensional.SIPL.JLucid.JIndexicalLucidParserTokenManager |
| gipsy.GIPC.intensional.SIPL.JOOIP.ast.visitor.DumpVisitor |
| gipsy.GIPC.intensional.SIPL.JOOIP.JavaCharStream |
| gipsy.GIPC.intensional.SIPL.JOOIP.JavaParser |
| gipsy.GIPC.intensional.SIPL.JOOIP.JavaParserTokenManager |
| gipsy.GIPC.intensional.SIPL.Lucx.LucxParser |
| gipsy.GIPC.intensional.SIPL.Lucx.LucxParserTokenManager |
| gipsy.GIPC.intensional.SIPL.ObjectiveLucid.ObjectiveGIPLParser |
| gipsy.GIPC.intensional.SIPL.ObjectiveLucid.ObjectiveGIPLParserTokenManager |
| gipsy.GIPC.intensional.SIPL.ObjectiveLucid.ObjectiveIndexicalLucidParser |
| gipsy.GIPC.intensional.SIPL.ObjectiveLucide.ObjectiveIndexicalLucidParserTokenmanager |
| gipsy.GIPC.Preprocessing.PreprocessorParser |
| gipsy.GIPC.Preprocessing.PreprocessorParserTokenManager |
| gipsy.GIPC.SemanticAnalyzer |
| gipsy.GIPCY.util.SimpleCharStream |
| gipsy.lang.GIPSYContext |
| gipsy.RIPE.editors.RunTimeGraphEditor.core.GlobalInstance |
| gipsy.RIPE.editors.RunTimeGraphEditor.ui.GIPSYGMTOperator |
| gipsy.tests.GIPC.intensional.SIPL.Lucx.SemanticTest.LucxSemanticAnalyzer |

*Figure 12: List of POOR Classes in GIPSY*

**Quality Checker Report using Logiscope**:

The following levels are available for Classes analysis:

- Factor level
- Criteria level
- Metric level

**Class Factor Level:**

In the following Pie Charts you will find for each factor which applies to classes:

The name of the factor.

The list of categories.

Firstly considering Gipsy's factor level, the maintainability is divided into excellent, good, fair and poor levels.

It can be seen that the poor classes are the least followed by fair then excellent and the maximum maintainability possibility can be seen in good level which accounts for 59% of the total. Similarly for DMARF, there is slight variation approximately 2% in the maintainability factor which can be seen in the diagram below.

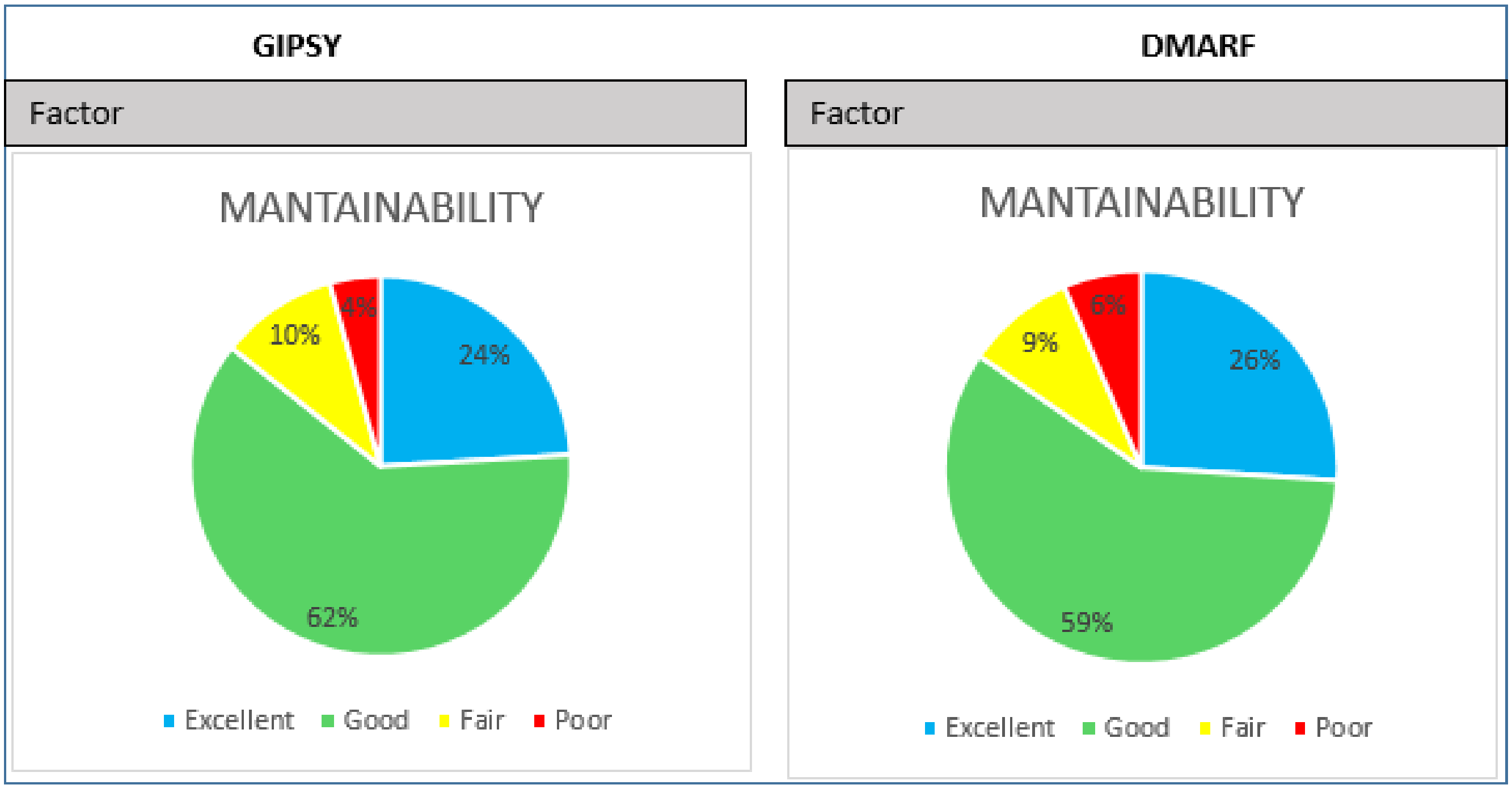

*Figure 13: Class Factor Level (GIPSY and DMARF)*

**Classes Criteria Level:**

In the following Pie Charts you will find for each criteria which applies to classes:

The name of the criteria.

The list of categories.

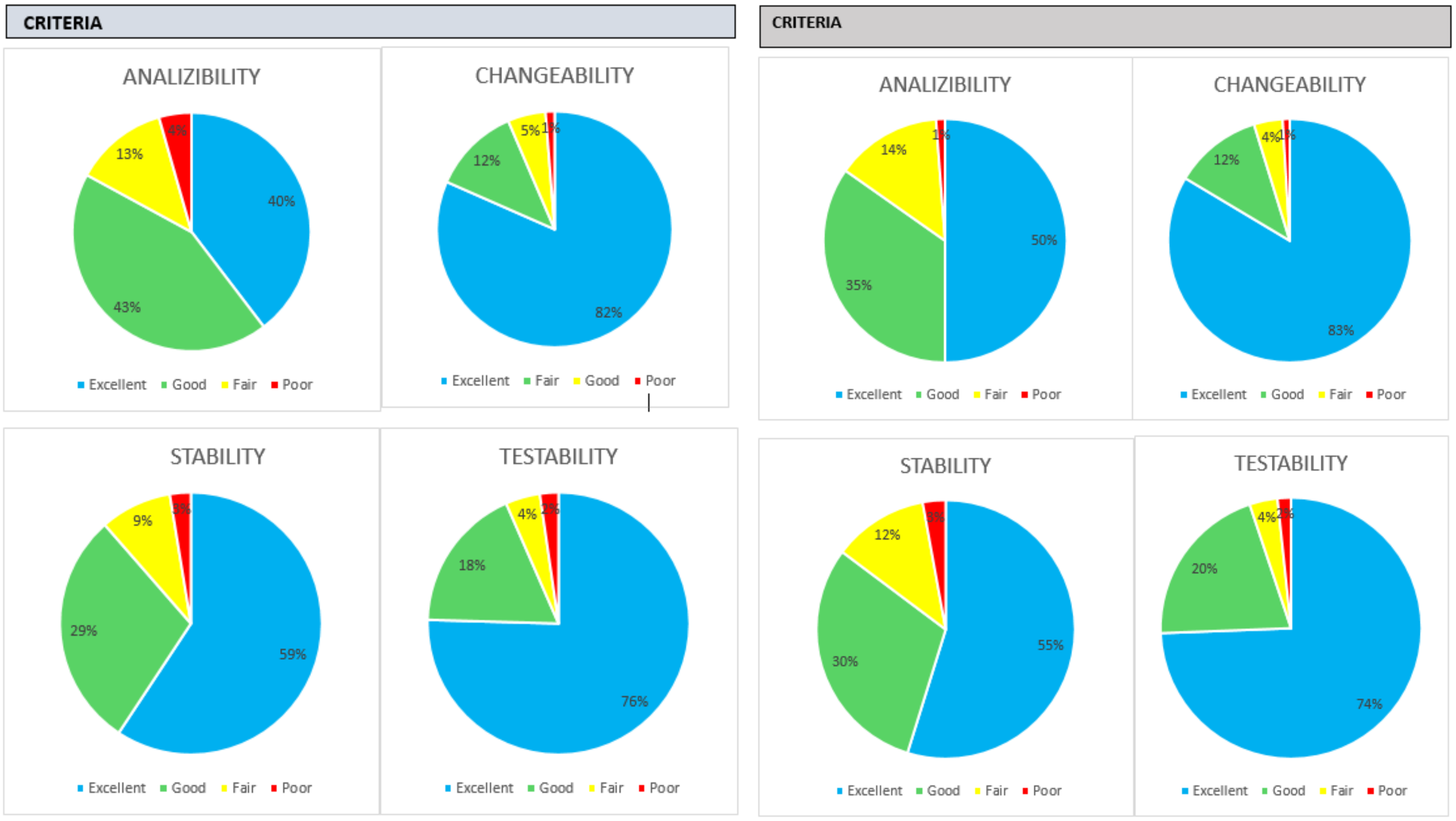

*Figure 14: Class Criteria Level (GIPSY and DMARF)*

**Class Metric Level:**

In the following table you will find for each metric which applies to classes:

The mnemonic and the name of the metric.

The min and max bounds for the metric.

The percentage of components out of bounds.

| Mnemonic | Metric Name | Min | Max | Out |
|---|---|---|---|---|
| cl_comf | Class comment rate | 0.20 | +00 | 40.24% |
| cl_comm | Number of lines of comment | -00 | +00 | 0.00% |
| cl_data | Total number of attributes | 0 | 7 | 17.84% |
| cl_data_publ | Number of public attributes | 0 | 0 | 28.66% |
| cl_func | Total number of methods | 0 | 25 | 5.64% |
| cl_func_publ | Number of public methods | 0 | 15 | 7.01% |
| cl_line | Number of lines | -00 | +00 | 0.00% |
| cl_stat | Number of statements | 0 | 100 | 11.43% |
| cl_wmc | Weighted Methods per Class | 0 | 60 | 6.10% |
| cu_cdused | Number of direct used classes | 0 | 10 | 21.65% |
| cu_cdusers | Number of direct users classes | 0 | 5 | 14.18% |
| in_bases | Number of base classes | 0 | 3 | 13.87% |
| in_noc | Number of children | 0 | 3 | 5.18% |

*Table 9: GIPSY Class Metric Level*

| Mnemonic | Metric Name | Min | Max | Out |
|---|---|---|---|---|
| cl_comf | Class comment rate | 0.20 | +00 | 8.10% |
| cl_comm | Number of lines of comment | -00 | +00 | 0.00% |
| cl_data | Total number of attributes | 0 | 7 | 9.52% |
| cl_data_publ | Number of public attributes | 0 | 0 | 27.62% |
| cl_func | Total number of methods | 0 | 25 | 4.76% |
| cl_func_publ | Number of public methods | 0 | 15 | 13.10% |
| cl_line | Number of lines | -00 | +00 | 0.00% |
| cl_stat | Number of statements | 0 | 100 | 7.86% |
| cl_wmc | Weighted Methods per Class | 0 | 60 | 3.10% |
| cu_cdused | Number of direct used classes | 0 | 10 | 24.76% |
| cu_cdusers | Number of direct users classes | 0 | 5 | 18.10% |
| in_bases | Number of base classes | 0 | 3 | 29.52% |
| in_noc | Number of children | 0 | 3 | 4.05% |

*Table 10: DMARF Class Metric Level*

## Kiviat Graph for Selected Classes (GIPSY):

**Kiviat Diagram for gipsy.RIPE.editors.RunTimeGraphEditor.core.GlobalInstance**

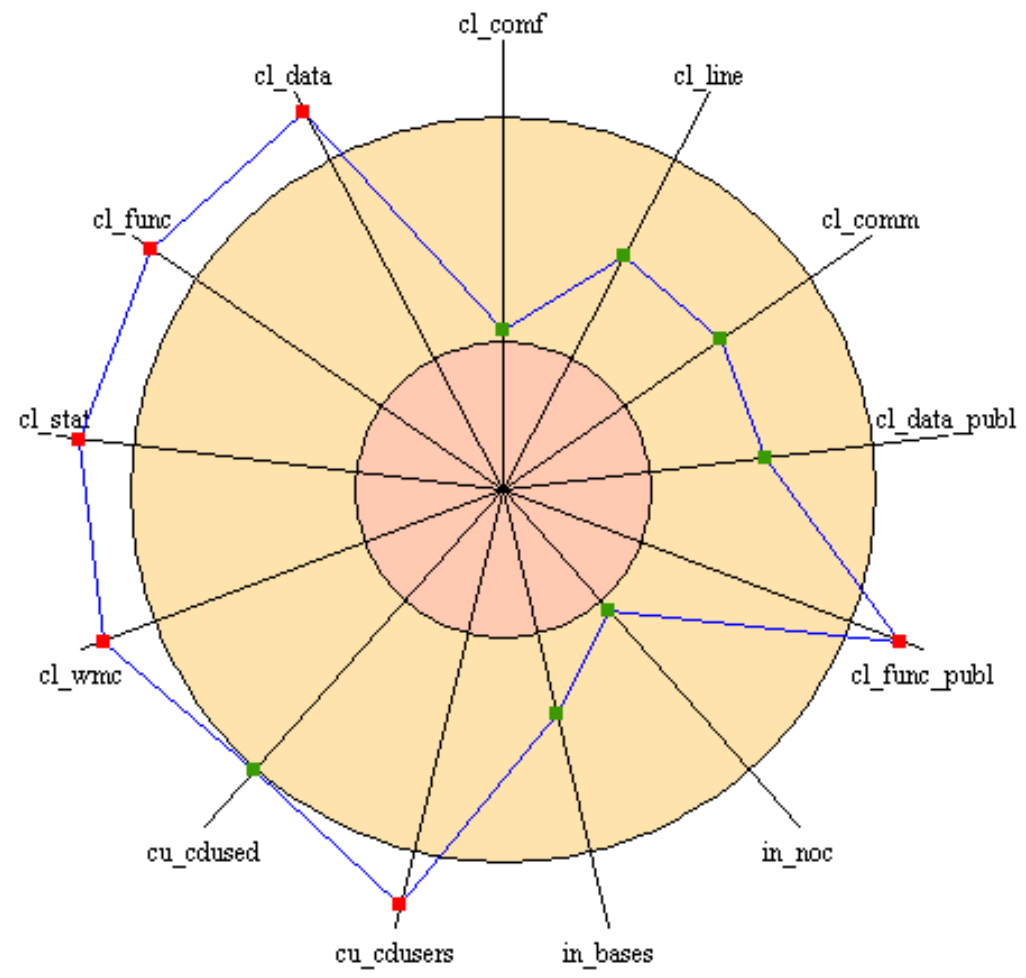

*Figure 15: Kiviat Diagram for GIPSY (i)*

**Kiviat Diagram for gipsy.GIPC.intensional.GenericTranslator.TranslationParser**

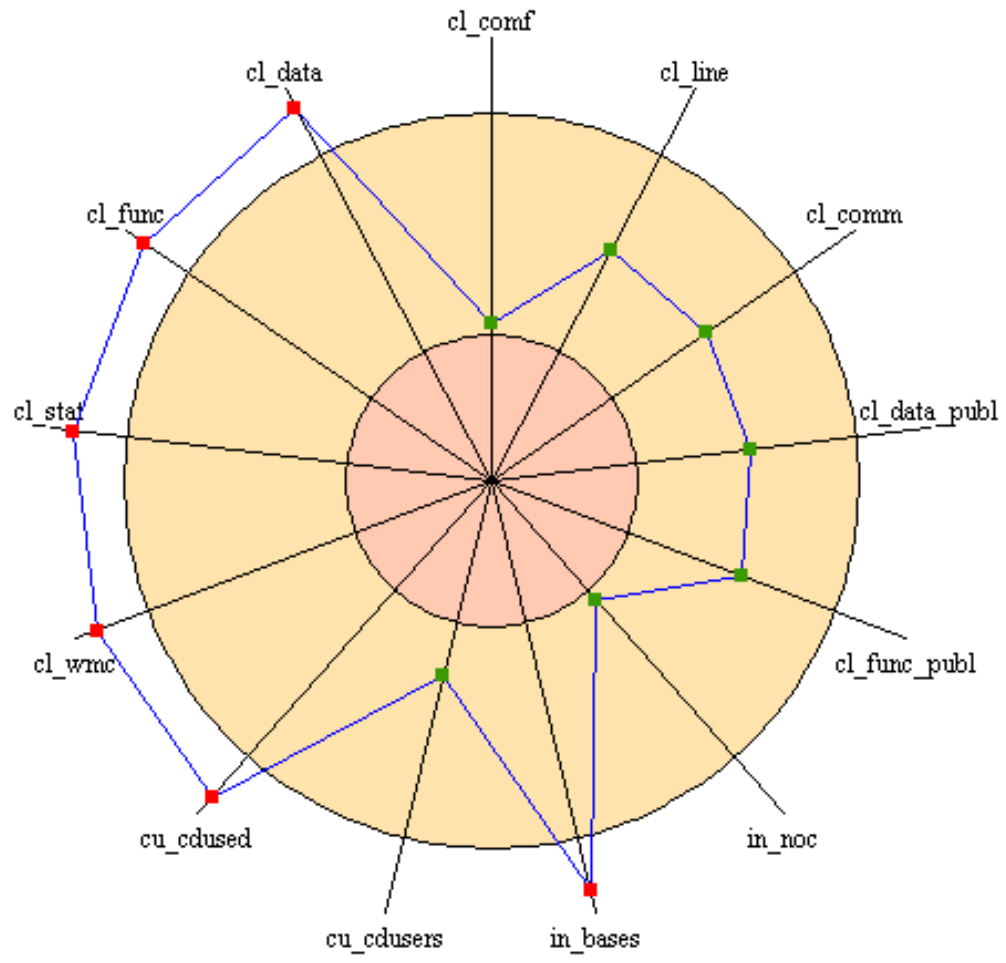

*Figure 16: Kiviat Diagram for GIPSY (i)*

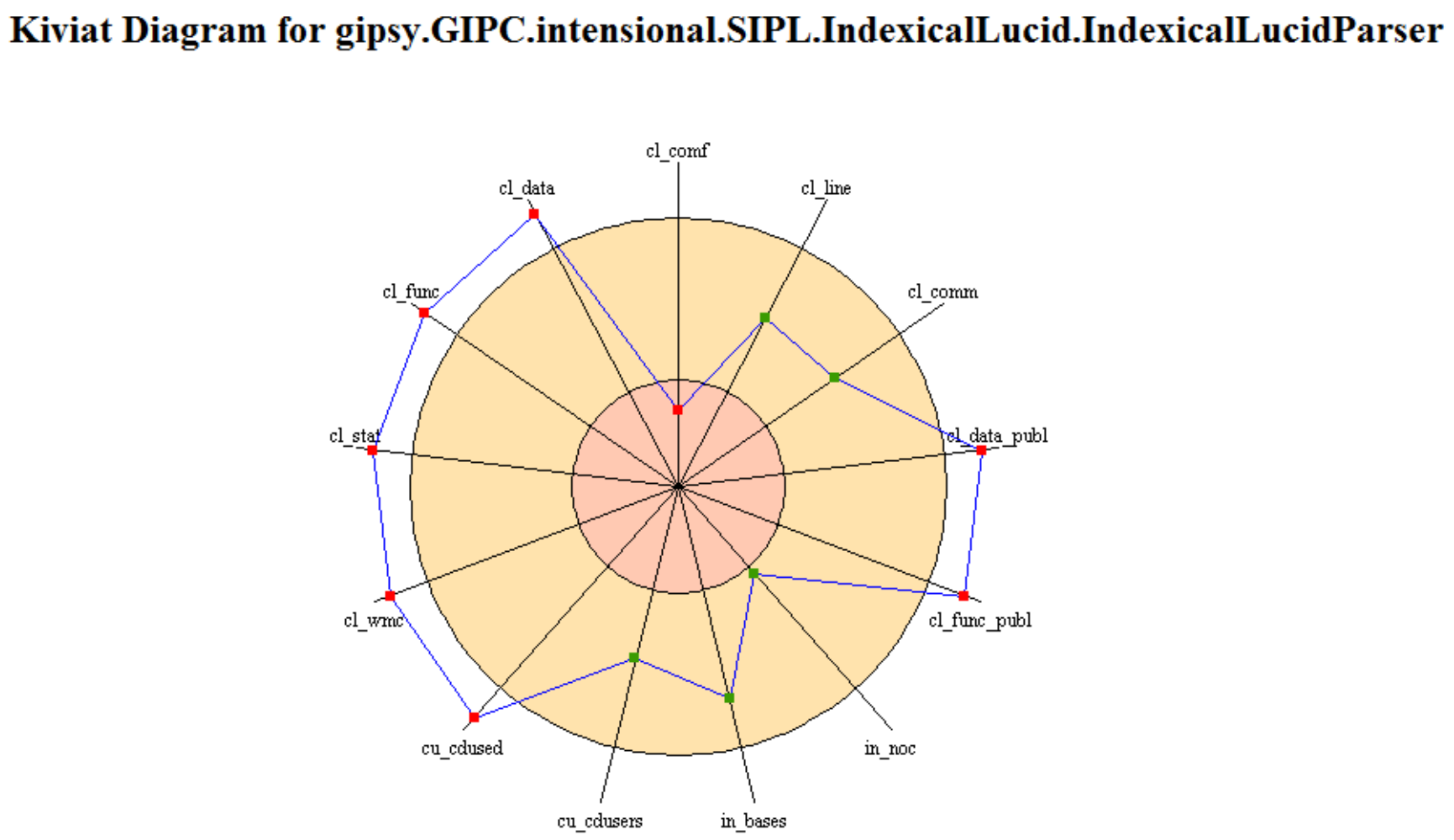

*Figure 17: Kiviat Diagram for GIPSY (iii)*

## Kiviat graphs for selected classes (DMARF):

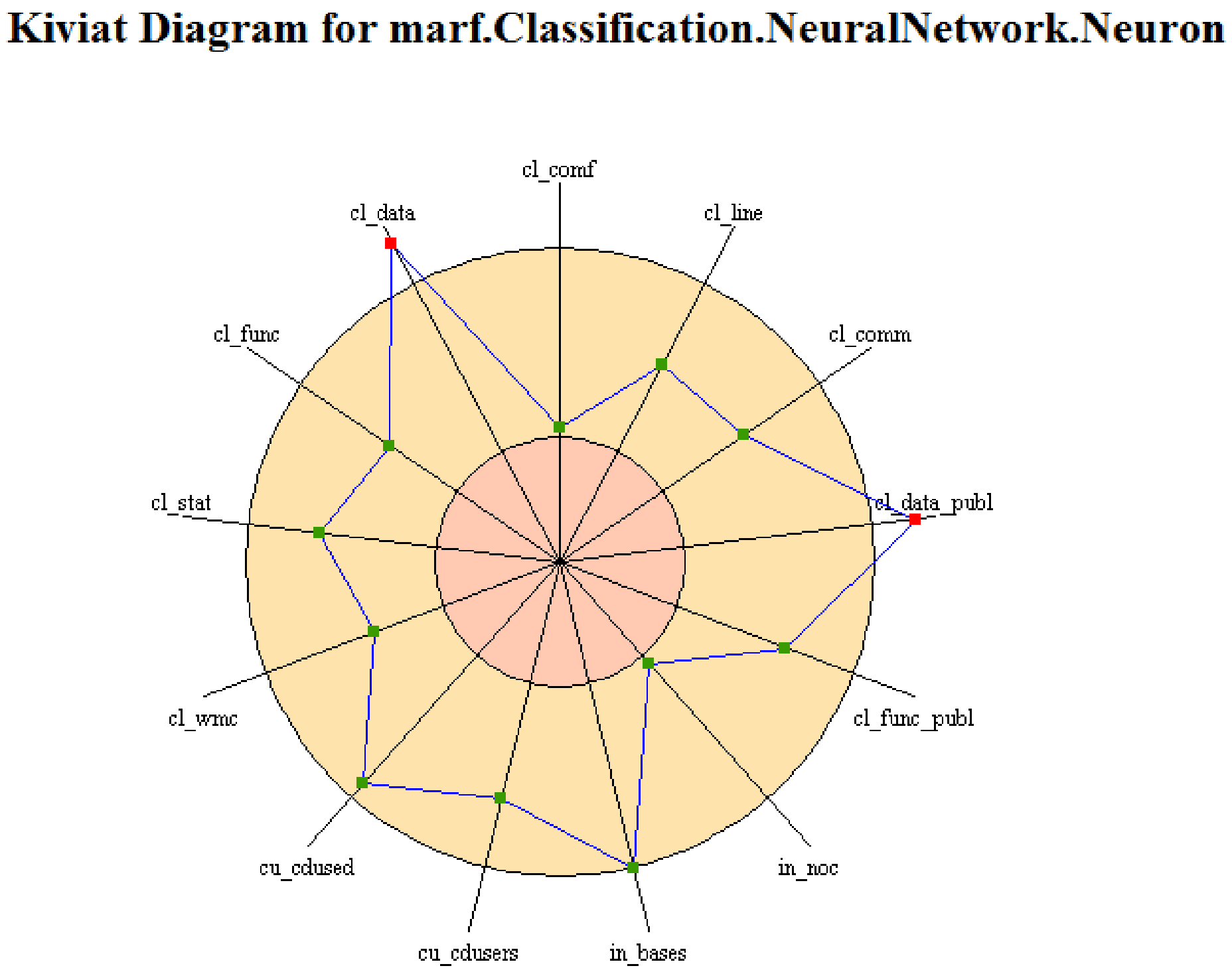

*Figure 18: Kiviat Diagram for DMARF (i)*

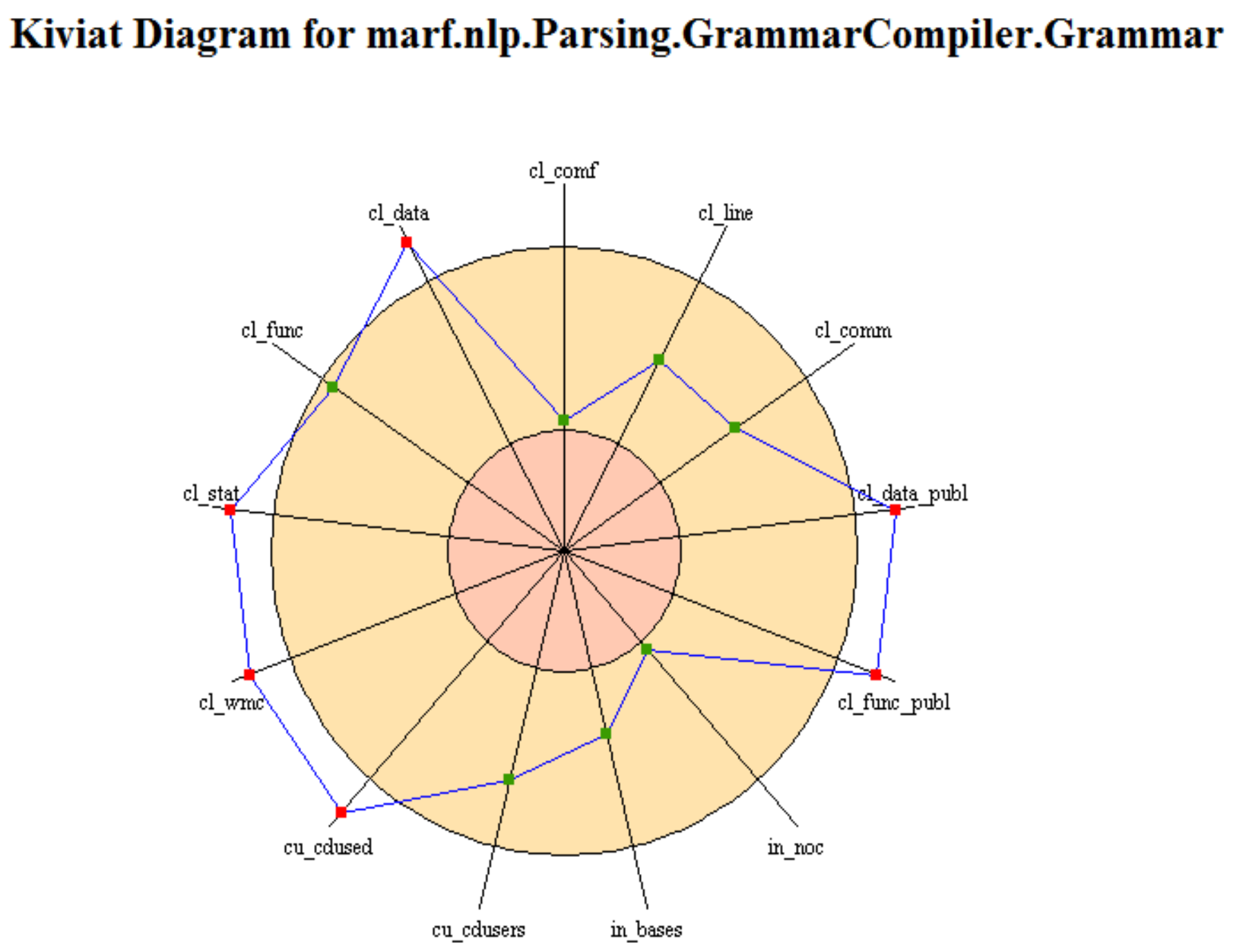

*Figure 19:Kiviat Diagram for DMARF (ii)*

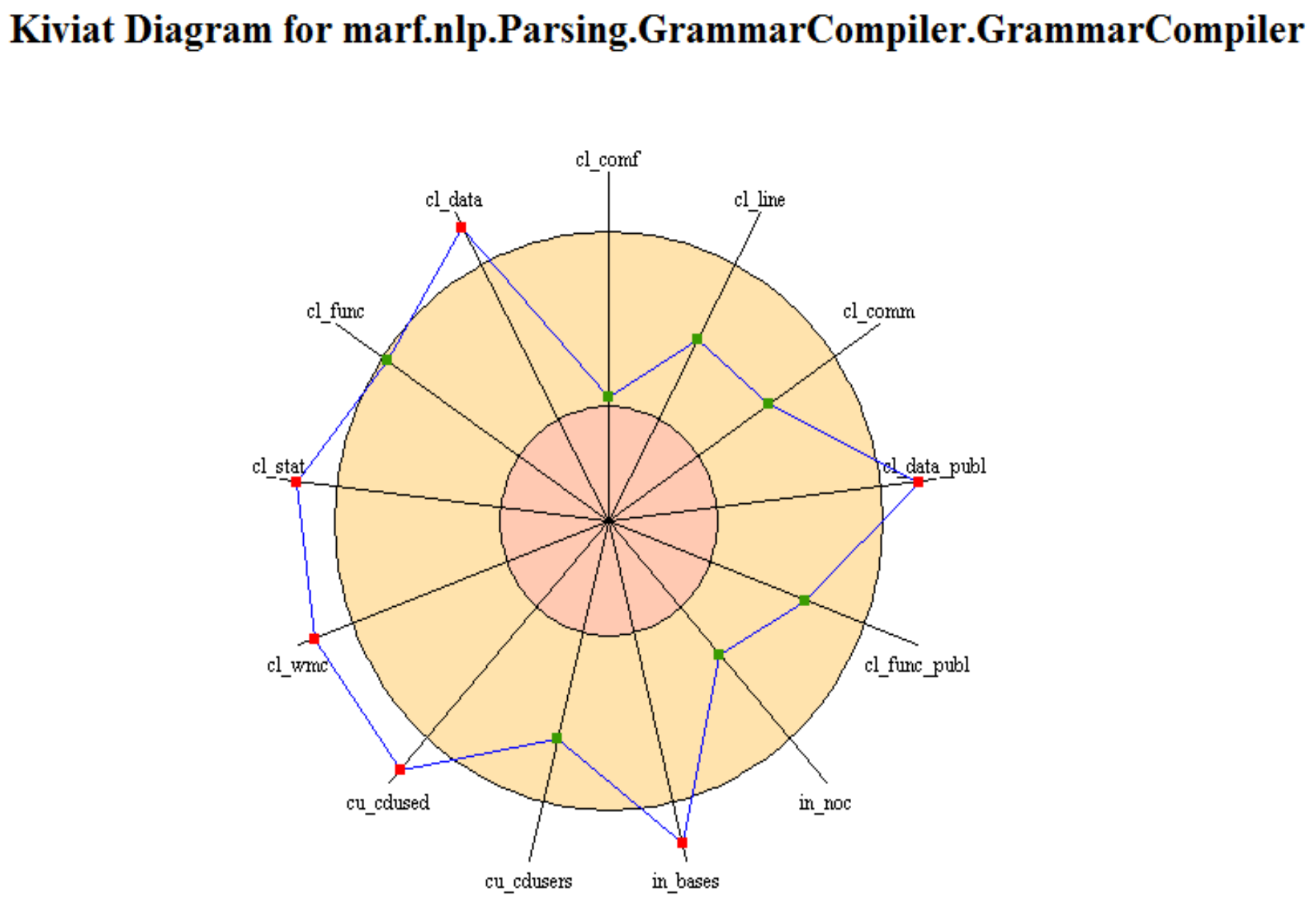

*Figure 20: Kiviat Diagram for DMARF (iii)*

Consider a class from the list of code smells:

**marf.nlp.Parsing.GrammarCompiler.Grammar**

In this class the method **ComputeFollowSets** method is not used in **marf.nlp.Parsing.GrammarCompiler.Grammar** so we can move this to another class named **marf.nlp.Parsing.GrammarCompiler.GrammarCompiler**

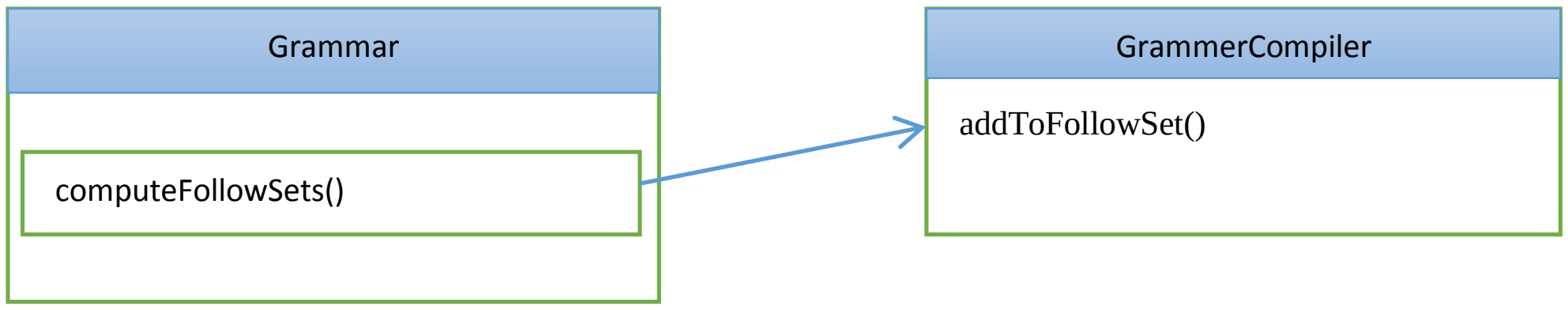

2. Planned Refactorings
DMARF & GIPSY

And we can rename the method to addToFollowSet(marf.nlp.Parsing.GrammarCompiler.Grammar) to reflect its new responsibility, so there is increase in cohesion and coupling is reduced since the related data is combined together the system is more efficient and the structure is improved.

Following are the refactoring which we will be implementing in PM4

- Pull Up Method

We have methods with identical results on subclasses, move them to superclass. The objective is to eliminate duplicate behavior. Although the same methods work fine within a code but they are building blocks of errors in future. So in order to avoid the risk we need to remove them.

- Removing clone (Duplicate) code

Finding the duplicate code from source code and removing it is one of the code refactoring technique. Using some existing duplicated code detection tools, we can find the existing code duplication so that we can remove it to make our code more efficient.

- Long Method

The longer the method gets more difficult is it to understand. In this we basically reduce the long lines of code (unnecessary code) reduce it to manageable and efficient code.

- Simplifying Conditional Expressions

If there exist some complex conditional expressions, they can be decomposed to make the code look simpler. It will help to improve readability of the code. Moreover it will make our code easy to understand.

- Make method calls Simpler

In such kind of refactoring we will be renaming methods who are having complex or ambiguous names and we will be creating extra objects to make our method calls simpler to increase the understandability and readability of the code.

While doing the GIPSY refactoring, we needed some test cases to verify that the behavior of system is not changed. We looked in to the gipsy.test directory to look for relevant test cases but we were unable to find one. So we created our own test cases and mentioned with the code in this document. They are named after the respective name of their class in which refactoring is performed. These test cases prove that refactoring improved the code but it didn't changed the behavior of the system. This is indeed one of the main concerns of refactoring.

3. Identification of Design Patterns
DMARF

1. Singleton Pattern

   **Definition:** Exactly one instance of a class is need. Object needs a global and single access point. Define a static method of the class that returns the Singleton pattern.

   **Reference:**
   - http://users.encs.concordia.ca/~s64711/lectures/14-design-patterns.pdf
   - http://uet.vnu.edu.vn/~chauttm/e-books/java/Head-First-Java-2nd-edition.pdf
   - http://www.oodesign.com/singleton-pattern.html

**Interacting Classes:** OptionFileLoader to itself.

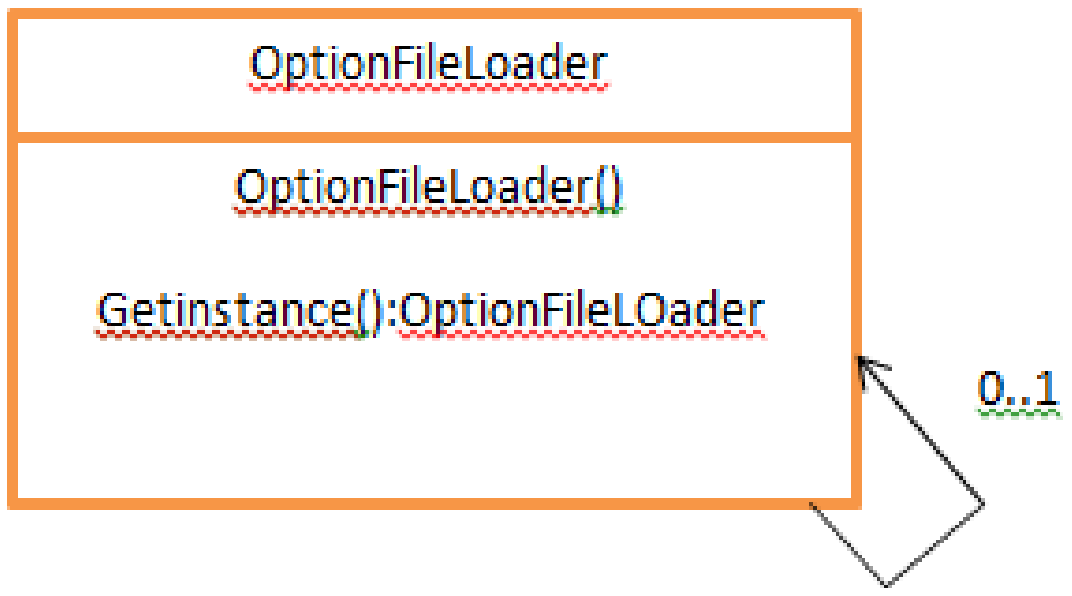

*Figure 21: Singleton Pattern*

**Reverse Engineering tool used:** Pattern4

**Source Code:**

```
public class OptionFileLoader
implements IOptionProvider
{
        /**
         * Singleton Instance.
         */
        protected static OptionFileLoader soOptionsLoaderInstance = null;
        .
        .
        .
        public static synchronized OptionFileLoader getInstance()
        {
                if(soOptionsLoaderInstance == null)
                {
                        soOptionsLoaderInstance = new OptionFileLoader();
                }
                return soOptionsLoaderInstance;
        }
}
```

2. Adapter Pattern

**Definition:** In the adapter pattern converts the original interface of a component in to another interface, through an intermediate adapter object.

**Reference:**

- http://users.encs.concordia.ca/~s64711/lectures/14-design-patterns.pdf
- http://uet.vnu.edu.vn/~chauttm/e-books/java/Head-First-Java-2nd-edition.pdf
- http://www.oodesign.com/adapter-pattern.html

**Interacting Classes:** FeatureExtraction class to StorageManager via IPreprocessing Interface.

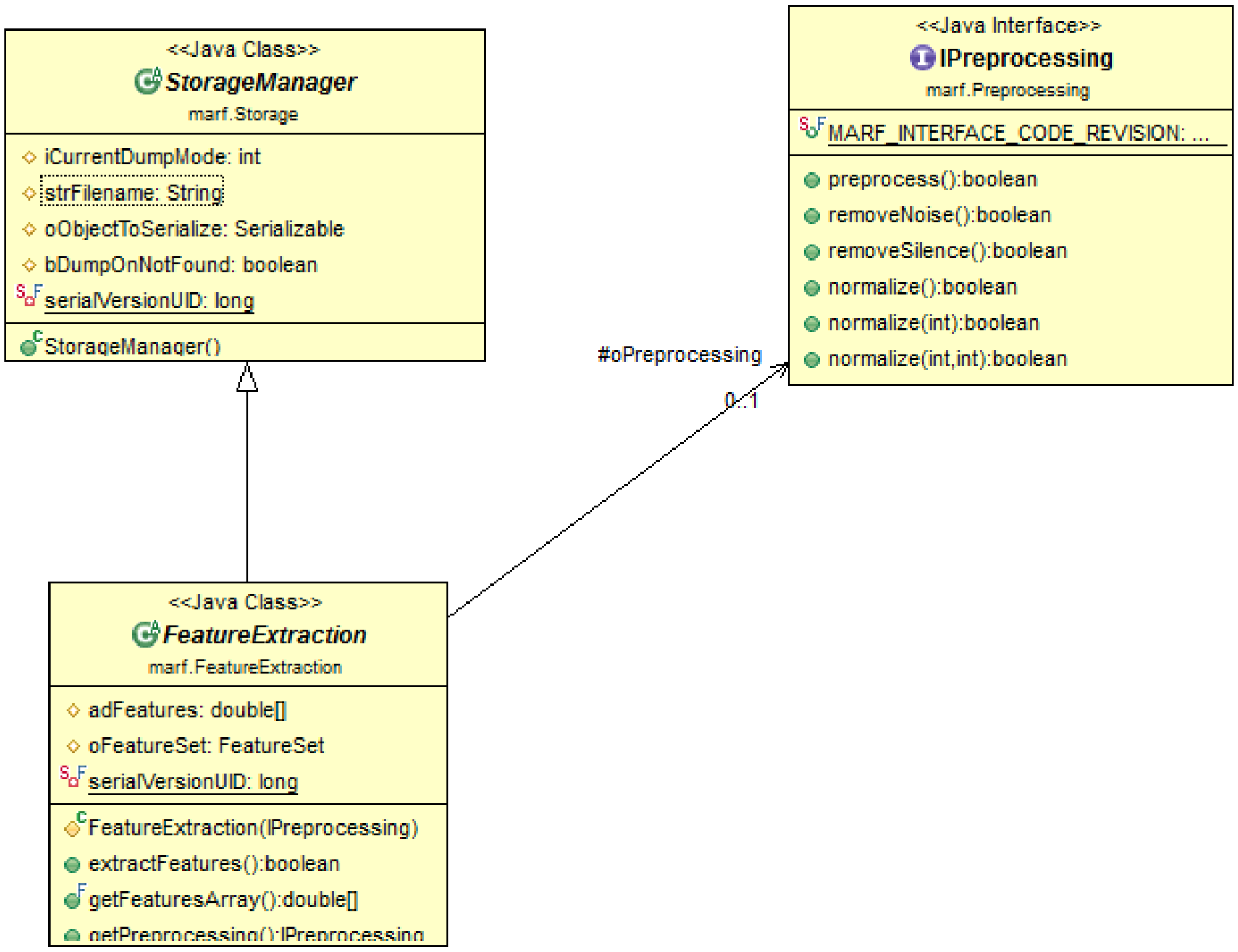

*Figure 22: Adapter Pattern*

**Reverse Engineering tool used:** Pattern4

**Source Code:**

```
public abstract class FeatureExtraction
extends StorageManager
implements IFeatureExtraction
{
	protected IPreprocessing oPreprocessing = null;
	.
	.
	.
	protected FeatureExtraction(IPreprocessing poPreprocessing)
	{
		/**
		 * adapter Instance.
		 */
		this.oPreprocessing = poPreprocessing;
		this.iCurrentDumpMode = DUMP_GZIP_BINARY;
		this.oObjectToSerialize = this.adFeatures;
	}
	.
	.
	.
}
```

3. State-Strategy Pattern

**Definition:** Defines an interface common to all supported algorithms. Context uses this interface to call the algorithm defined by a ConcreteStrategy.

**Reference:**

- http://users.encs.concordia.ca/~s64711/lectures/14-design-patterns.pdf
- http://uet.vnu.edu.vn/~chauttm/e-books/java/Head-First-Java-2nd-edition.pdf
- http://www.oodesign.com/strategy-pattern.html

**Interacting Classes:** Preprocessing and ipreprocessing

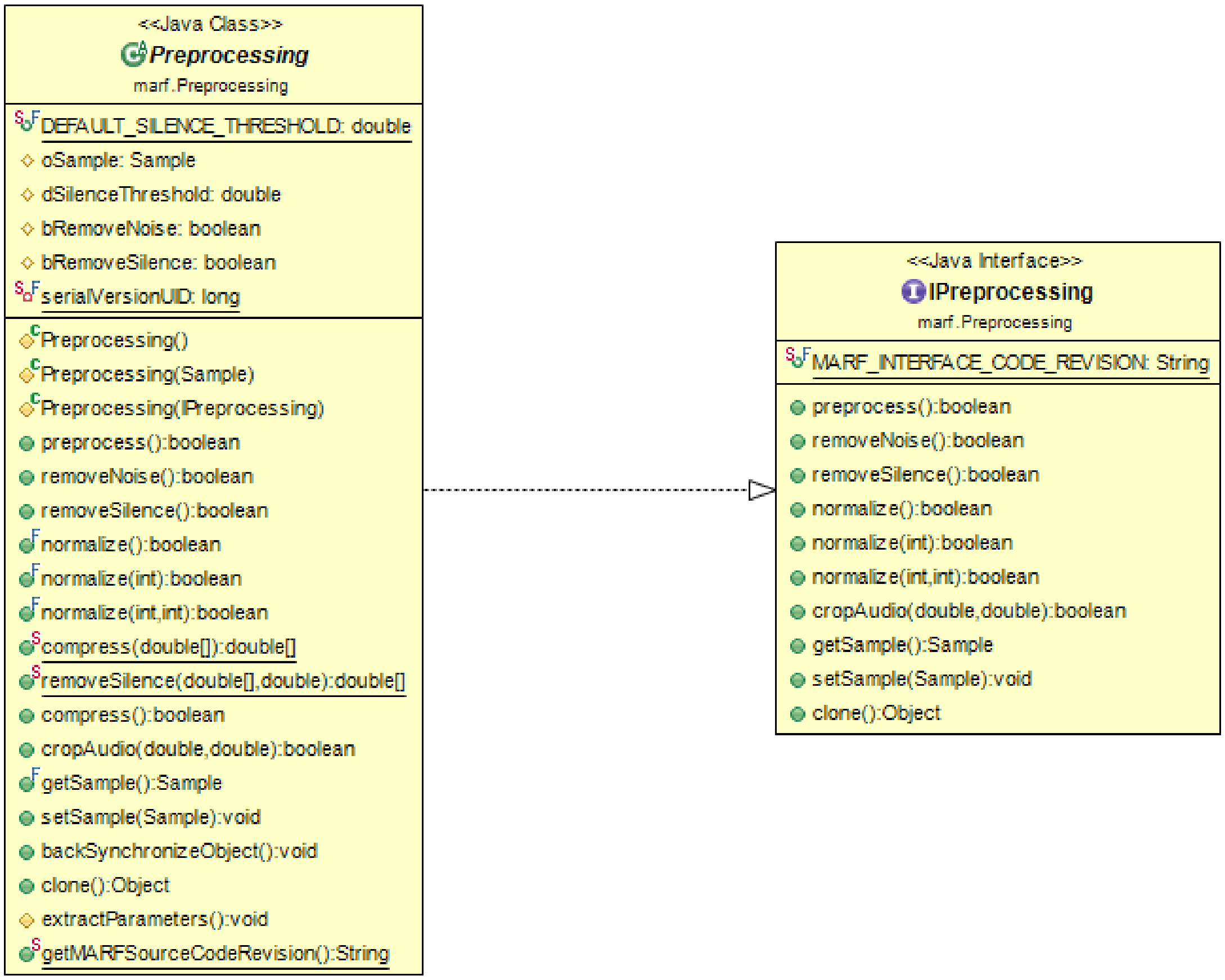

*Figure 23: state-Strategy Pattern*

**Reverse Engineering tool used:** Pattern4

**Source Code:**

```
public abstract class Preprocessing
extends StorageManager
implements IPreprocessing
	{
	.
	.
	.
	protected Preprocessing(IPreprocessing poPreprocessing)
		throws PreprocessingException
		{
			if(poPreprocessing == null)
			{
				throw new IllegalArgumentException("Preprocessing parameter cannot be null.");
			}
			boolean bChanged = poPreprocessing.preprocess();
			if(bChanged == false)
			{
				Debug.debug
				(
					"WARNING: " +
					poPreprocessing.getClass().getName() +
					".preprocess() returned false."
				);
			}
```

```
        this.oObjectToSerialize =
this.oSample =
poPreprocessing.getSample();

        extractParameters();
            }
.
.
.
}
```

4. Prototype Pattern

**Definition:**

- specifying the kind of objects to create using a prototypical instance
- creating new objects by copying this prototype

**Reference:**

- http://www.oodesign.com/prototype-pattern.html
- http://en.wikipedia.org/wiki/Prototype_pattern

**Interacting Classes:** Classification class with clone():java.Object.lang class

**Reverse Engineering Tool used:** Pattern4

**Source Code:**

```
/**
    * prototype instance
*/
public Object clone()
        {
                Classification oClone =
(Classification)super.clone();
                oClone.oResultSet =
(ResultSet)this.oResultSet.clone();
                oClone.oTrainingSet =
(TrainingSet)this.oTrainingSet.clone();
                oClone.oFeatureExtraction =
this.oFeatureExtraction;
                return oClone;
        }
```

| Sl. No. | Source | Recognized Pattern | Interacting Class(s) | Reverse Engineering Tool Used |
|---|---|---|---|---|
| 1 | DMARF | Singleton | OptionFileLoader | Pattern4 |
| 2 | DMARF | Adapter | FeatureExtraction class to StorageManager via IPreprocessing | Pattern4 |
| 3 | DMARF | State Strategy | Preprocessing and ipreprocessing | Pattern4 |
| 4 | DMARF | Prototype | Classification class with clone():java.Object.lang class | Pattern4 |

*Table 11: DMARF Design Pattern Summary Table*

GIPSY

1. Factory Pattern

   **Definition:** Define an interface for creating an object, but let subclasses decide which class to instantiate. Factory lets a class defer instantiation to subclasses. It basically creates a pure fabrication called a factory that handles the creation.

   **Reference:**

   - http://users.encs.concordia.ca/~s64711/lectures/14-design-patterns.pdf
   - http://uet.vnu.edu.vn/~chauttm/e-books/java/Head-First-Java-2nd-edition.pdf
   - http://www.oodesign.com/factory-pattern.html

**Interacting Classes:** TierFactoryDGTFactoryDSTFactoryDWTFactor

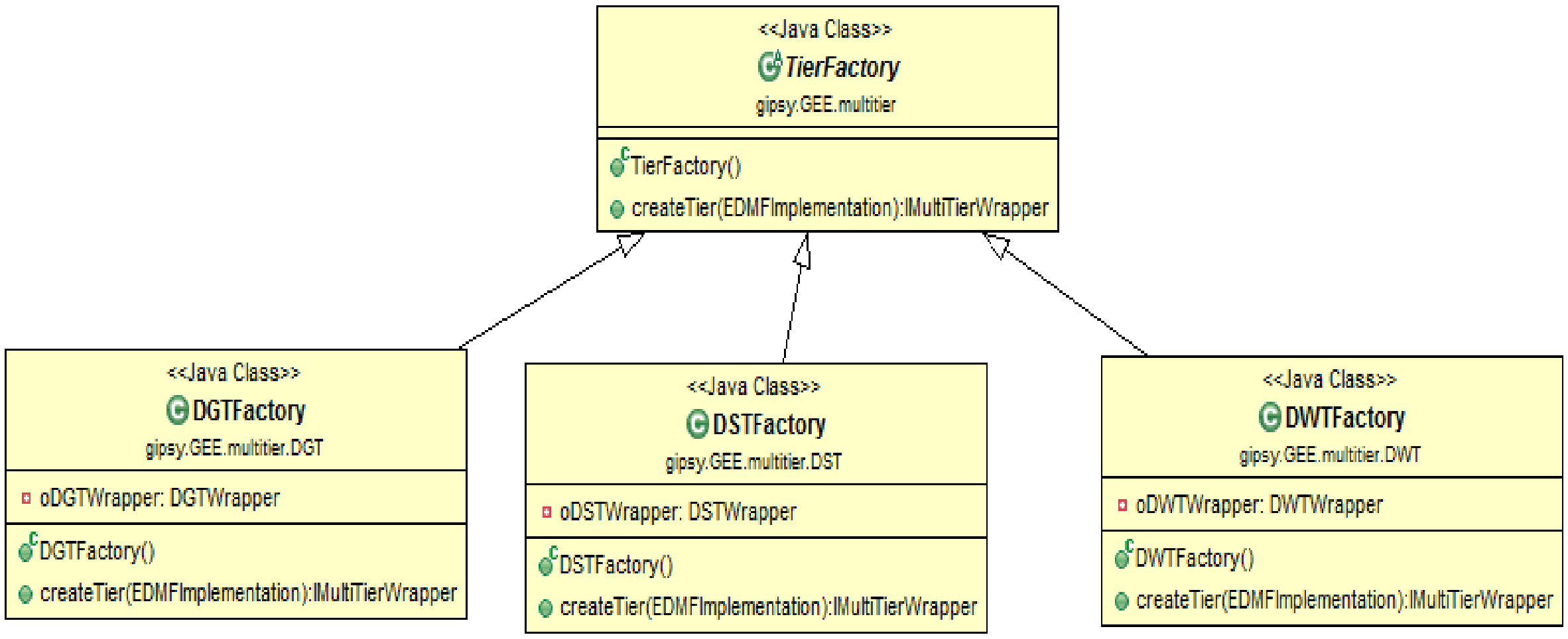

*Figure 24: Factory Pattern*

Tier Factory is the base class where as DGT, DST, and DWT are the sub classes. As mentioned DGT DST and DWT will decide which classes to instantiate. And the process will be hidden from the base class that is the Tier Factory.

**Reverse Engineering tool used:** Pattern4

**Source Code:**

```
public abstract class TierFactory
{
        public IMultiTierWrapper createTier(EDMFImplementation poDMFImp)
        throws MultiTierException
        {
                return null;
        }
.
.
.
        /**
         * Factory pattern instance
         */
public IMultiTierWrapper createTier(String pstrType)
        {
                IMultiTierWrapper oTierWrapper = null;
                if(pstrType.equals("DST"))
                {
                        oTierWrapper = new DSTWrapper();
                }
                else if(pstrType.equals("DWT"))
                {
                        oTierWrapper = new DWTWrapper();
                }
                else if(pstrType.equals("DGT"))
                {
                        oTierWrapper = new DGTWrapper();
                }

                return oTierWrapper;
        }
    }
```

2. Observer Pattern

**Definition:** Define a one-to-many dependency between objects so that when one object changes state, all its dependents are notified and updated automatically. Object should be able to notify others that may not be known from the beginning.

**Reference:**

- http://users.encs.concordia.ca/~s64711/lectures/14-design-patterns.pdf
- http://uet.vnu.edu.vn/~chauttm/e-books/java/Head-First-Java-2nd-edition.pdf
- http://www.oodesign.com/observer-pattern.html

**Interacting Classes:** JJForesnsicLucidParserState to Node class.

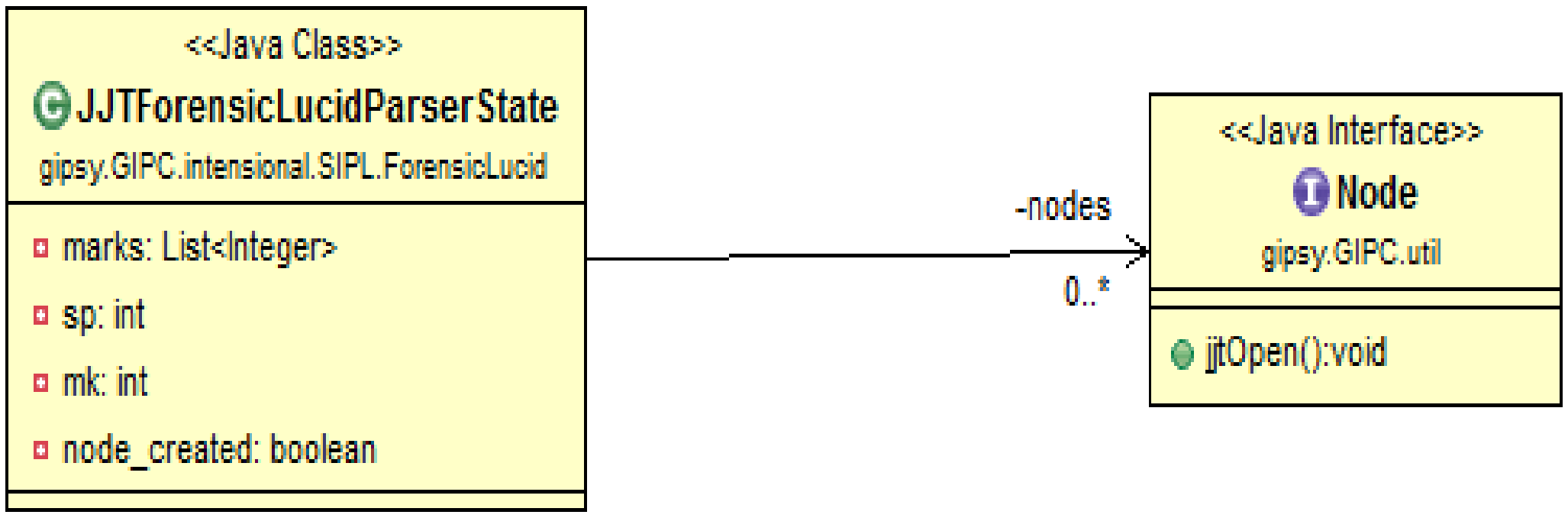

*Figure 25: Observer Pattern*

**Reverse Engineering tool used:** Pattern4

**Source Code:**

```
import gipsy.GIPC.util.Node;

public class JJTForensicLucidParserState {

private java.util.List<Node> nodes;

private java.util.List<Integer> marks;

.

.

.

private int sp;        // number of nodes on stack

private int mk;        // current mark

private boolean node_created;

public JJTForensicLucidParserState() {

nodes = new java.util.ArrayList<Node>();

marks = new java.util.ArrayList<Integer>();

sp = 0;

mk = 0;

}
```

3. Decorator Pattern

**Definition:** The intent of this pattern is to add additional responsibilities dynamically to an object.

**Reference:**

- http://users.encs.concordia.ca/~s64711/lectures/14-design-patterns.pdf
- http://uet.vnu.edu.vn/~chauttm/e-books/java/Head-First-Java-2nd-edition.pdf
- http://www.oodesign.com/decorator-pattern.html

**Interacting Classes:** IDemandWorkerDemandWorkerMARFCATDWTAPP

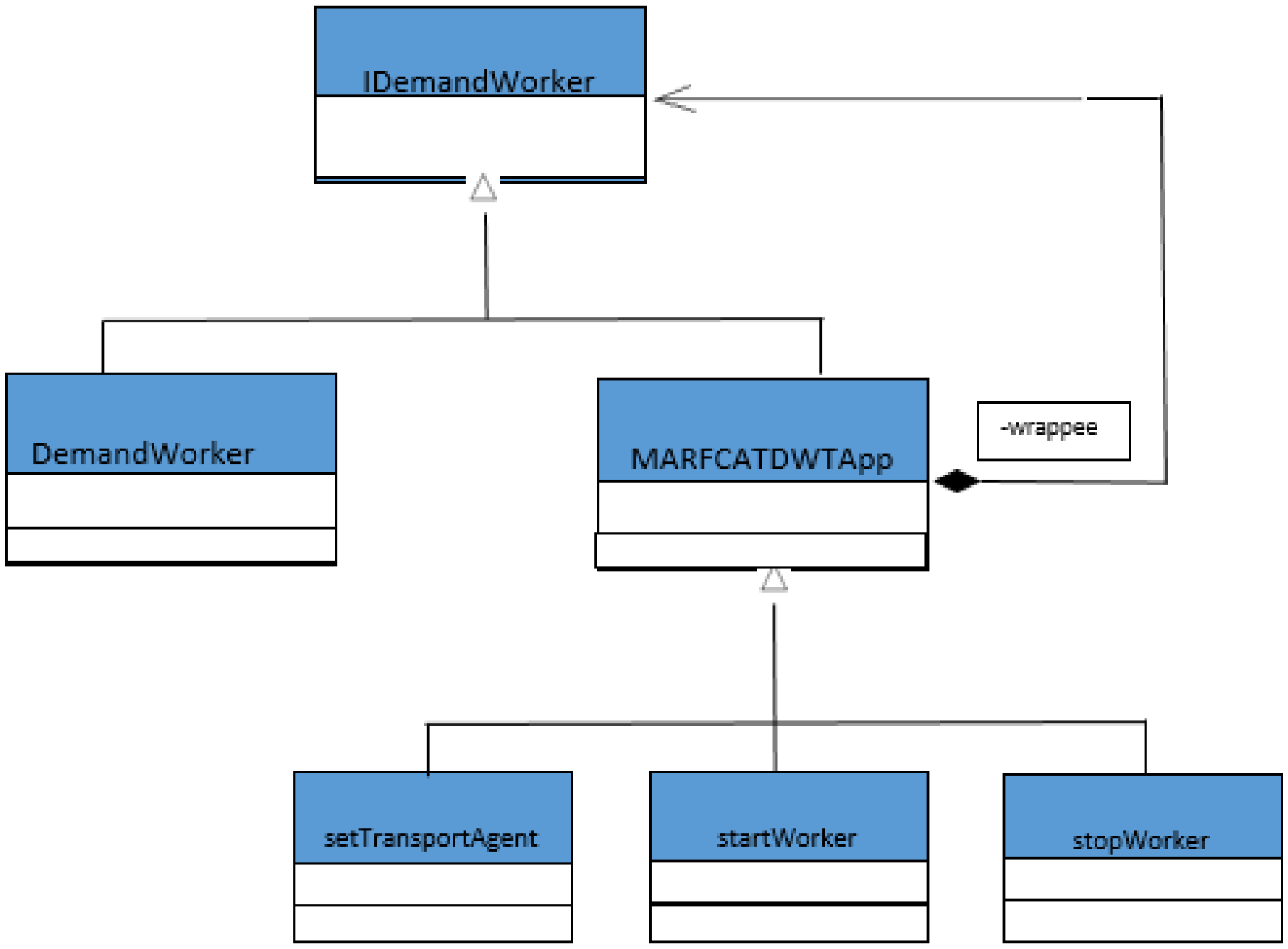

*Figure 26: Decorator Pattern*

**Reverse Engineering tool used:** Pattern4

**Source Code:**

**(Interface)**

```
public interface IDemandWorker
extends Runnable
{
	void setTransportAgent(EDMFImplementation poDMFImp);
	void setTransportAgent(ITransportAgent poTA);
	void setTAExceptionHandler(TAExceptionHandler poTAExceptionHandler);
	void startWorker();
	void stopWorker();
}
```

**(CoreFunctionality)**

```
public class DemandWorker
implements IDemandWorker
{
	protected ITransportAgent oTA;
	protected LocalDemandStore oLocalDemandStore;
}
```

4. State-Strategy Pattern

**Definition:** Defines an interface common to all supported algorithms. Context uses this interface to call the algorithm defined by a ConcreteStrategy.

**Reference:**

- http://users.encs.concordia.ca/~s64711/lectures/14-design-patterns.pdf
- http://uet.vnu.edu.vn/~chauttm/e-books/java/Head-First-Java-2nd-edition.pdf
- http://www.oodesign.com/strategy-pattern.html

**Interacting Classes:** GEE interacting with IMultiTierWrapper class for finding number of tier wrappers available in that instance.

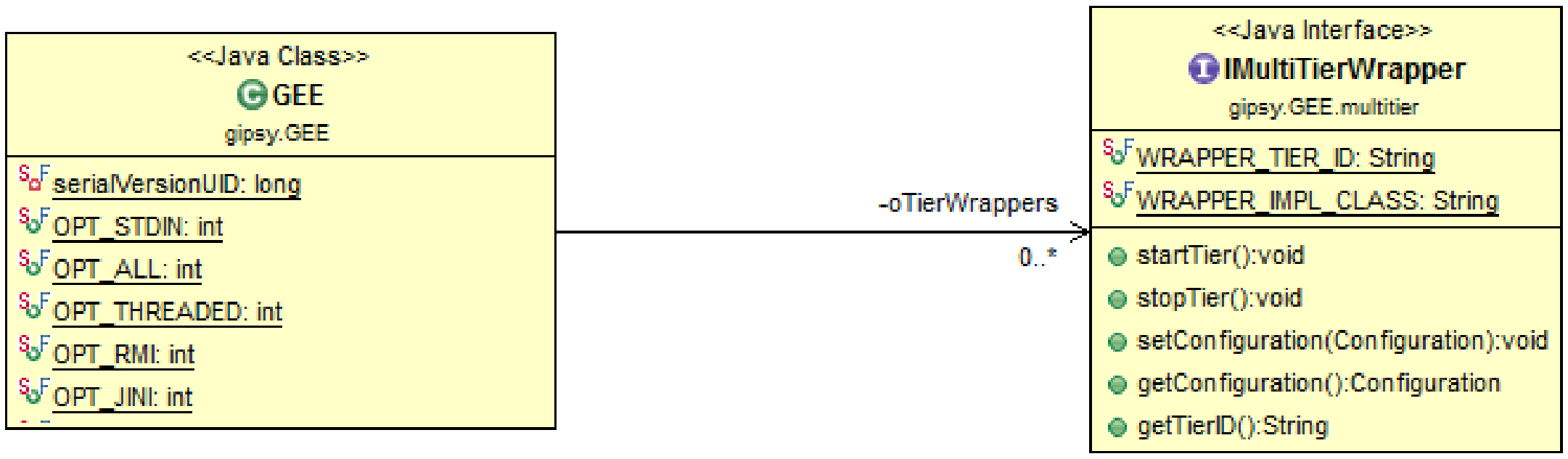

*Figure 27: State-Strategy Pattern*

**Reverse Engineering tool used:** Pattern4

**Source Code:**

```
private ArrayList<IMultiTierWrapper>
oTierWrappers = null;
.
.
.
public void startServices()
        throws GEEException
        {
            Debug.debug("GEE:
startServices() is somewhat implemented...");
            this.oTierWrappers = new
ArrayList<IMultiTierWrapper>();
.
.
.
        }
```

| Sl. No. | Source | Recognized Pattern | Interacting Class(s) | Reverse Engineering Tool Used |
|---|---|---|---|---|
| 1 | GIPSY | Factory | TierFactoryDGTFactoryDSTFactory<br>DWTFactor | Pattern4 |
| 2 | GIPSY | Observer | JJForesnsicLucidParserState to Node class. | Pattern4 |
| 3 | GIPSY | Decorater | IDemandWorkerDemandWorker<br>MARFCATDWTAPP | Pattern4 |
| 4 | GIPSY | State-Strategy | GEE interacting with IMultiTierWrapper class for finding number of tier wrappers available in that instance. | Pattern4 |

*Table 12: GIPSY Design Pattern summary Table*

# VI. Implementation

## A. Refactoring

### DMARF

a) Class: Neuron.java

Method: void train ()

In this class the method train has implemented Switch loop but there is a lot of code inside one of the switch case (Hidden Case).Thus the process of Extraction is used by providing new method called calculateDSum()

Before:

```
public final void train(final double pdExpected, final double pdAlpha, final double pdBeta)
{
    switch(this.iType)
    {
    case HIDDEN:
        {
            double dSum = 0.0;
            for(int i = 0; i < this.oOutputs.size(); i++)
            {
                dSum +=
    ((Neuron)this.oOutputs.get(i)).dDelta *
    ((Neuron)this.oOutputs.get(i)).getWeight(this);
            }
            this.dDelta = this.dResult * (1.0 - this.dResult) * dSum;
            break;
        }
    }
```

After:

```
    case HIDDEN:
        {
            double dSum = calculateDSum();

            this.dDelta = this.dResult * (1.0 - this.dResult) * dSum;
            break;
        }
    private double calculateDSum()
    {
        double dSum = 0.0;
        for(int i = 0; i < this.oOutputs.size(); i++)
        {
            dSum +=
    ((Neuron)this.oOutputs.get(i)).dDelta *
    ((Neuron)this.oOutputs.get(i)).getWeight(this);
        }
        return dSum;
    }
```

b) Class: NeuralNetwork.java

Method: void createLinks(Node poNode)

In this class named NeuralNetwork.java there is one long method createlinks().There was a huge amount of complex code which was making the system less cohesive. Due to this complex method the overall structure of the class was less cohesive.

We have refactored this long method by extracting relational code into one method. There are four such methods namely processLayer, processNeuron, createInputLink, createOutputLink.

Before:

```
private final void createLinks(Node poNode)
    throws ClassificationException
    {
        int iType = poNode.getNodeType();
        String strName;
        if(iType == Node.ELEMENT_NODE)
        {
            strName = poNode.getNodeName();
            NamedNodeMap oAtts = poNode.getAttributes();
            if(strName.equals("layer"))
            {
                .
                .
                .
            }
            else if(strName.equals("neuron"))
            {
                .
                .
                .
```

```
                            }
                            else
if(strName.equals("input"))
                            {
                                    .
                                    .
                                    .
                            }
                            else
if(strName.equals("output"))
                            {
                                    .
                                    .
                                    .
                            }
                    // Recurse for children if any
                    for
                    ( )
                    {
                            createLinks(oChild);
                    }
    }
```

After:

```
private final void createLinks(Node poNode)
          throws ClassificationException
          {
                    int iType = poNode.getNodeType();
                    String strName;
                    if(iType == Node.ELEMENT_NODE)
                    {
                            strName                    =
poNode.getNodeName();
                            NamedNodeMap    oAtts    =
poNode.getAttributes();
                            if(strName.equals("layer"))
                            {

          processLayer(oAtts);
                            }
                            else
if(strName.equals("neuron"))
                            {

          processNeuron(oAtts);
                            }
                            else
if(strName.equals("input"))
                            {

          createInputLink(oAtts);
                            }
                            else
if(strName.equals("output"))
                            {

          createOutputLink(oAtts);
                            }
                    }
                    // Recurse for children if any
                    for
                    ()
                    {
                            createLinks(oChild);
                    }
    }
          private  void  createOutputLink(NamedNodeMap
oAtts)
                            throws
ClassificationException
                    {
                    }
          private   void   createInputLink(NamedNodeMap
oAtts)
                            throws
ClassificationException
                    {
                    }
          private   void   processNeuron(NamedNodeMap
oAtts)
                    {
                    }
          private    void    processLayer(NamedNodeMap
oAtts)
                    {
                    }
```

GIPSY

a) Class: TranslationParser.java

Method: Parse()

Here in Parse method the if condition is making the code very complex so we have introduced a medial condition method in which we are using switch cases, Thus, it can be easily seen that cohesion within the methods is increased and complexity is reduced.

Before:

```
class TranslationParser
implements T, G, GIPLParserConstants, GIPLParserTreeConstants
{
}
public Hashtable Parse() throws IOException
{
    .
    .
    .
        if (strTemp.charAt(1) == '1')
        {
            DoCond();
        }
        else if (strTemp.charAt(1) == '2')
        {
            DoThen();
        }
        else if (strTemp.charAt(1) == '3')
        {
            DoElse();
        }
        else if (strTemp.charAt(1) == '4')
        {
            DoAt();
        }
        else if (strTemp.charAt(1) == '5')
        {
            DoEqual();
        }
        else if (strTemp.charAt(1) == '6')
        {
            DoBrace();
        }
        else if (strTemp.charAt(1) == '7')
        {
            DoNotBrace();
        }
}
```

**After:**

```
public Hashtable Parse() throws IOException
{
    .
    .
    .
    medialConditions(choice)

}

private void medialConditions() {
    switch(choice)
    {
    case '1': DoCond();
        break;
    case '2': DoThen();
        break;
    case '3': DoElse();
        break;
    case '4': DoAt();
        break;
    case '5': DoEqual();
        break;
    case '6': DoBrace();
        break;
    case '7': DoNotBrace();
        break;
    default: System.out.println("Not Valid value");
    }
}
```

b) Class: GlobalInstance.java

Method: isItemExists()

Before:

```
public boolean isItemExists(Object pItemToAdd)
  {
    if ((pItemToAdd instanceof GIPSYPhysicalNode)
        && (this.oPhysicalNodesList.contains(pItemToAdd)))
      return true;
    if ((pItemToAdd instanceof GIPSYInstance)
        && (this.oGIPSYInstanceList.contains(pItemToAdd)))
      return true;
    if ((pItemToAdd instanceof GIPSYTier)
        && (this.oGIPSYTiersList.contains(pItemToAdd)))
      return true;
    if ((pItemToAdd instanceof NodeConnection)
```

```java
        && (this.oNodeConnectionsList.contains(pItemToAdd)))
            return true;

        return false;
    }
```

After:

```java
    public boolean isItemExists(Object pItemToAdd)
    {
        if (((pItemToAdd instanceof GIPSYPhysicalNode)
            && (this.oPhysicalNodesList.contains(pItemToAdd)))
            || ((pItemToAdd instanceof GIPSYInstance)
             && (this.oGIPSYInstanceList.contains(pItemToAdd)))
            || ((pItemToAdd instanceof GIPSYTier)
             && (this.oGIPSYTiersList.contains(pItemToAdd)))
            || ((pItemToAdd instanceof NodeConnection)
             && (this.oNodeConnectionsList.contains(pItemToAdd))))
                                        {
                                            return true;
                                        }
        return false;
    }
```

c) Class: Configuration.java

Method: equals(Object pObject)

Before:

```java
public class Configuration
implements Serializable
{
public boolean equals(Object pObject)
        {
                if(pObject instanceof Configuration)
                {
                        return this.oConfigurationSettings.equals(((Configuration)pObject).oConfigurationSettings);
                }
                else
                {
                        return false;
                }
        }
}
```

After:

```java
public class Configuration
implements Serializable
{
public boolean equals(Object pObject)
        {
                boolean isValidConfig = false;

                if(pObject instanceof Configuration)
                {
                        Object setting = ((Configuration)pObject).oConfigurationSettings;
                        isValidConfig = this.oConfigurationSettings.equals(setting);

                }

                return isValidConfig;
}
```

## B. ChangeSets and Diffs

### 1. Neuron.java

**<u>Change 0/2: Refactor void *train()* into two methods of Class *Neuron.java*</u>**

The switch case of class '*Neuron.java'* has many line of codes within the particular case '*Hidden*'. Using Extraction method technique of Refactoring, we created a new method called '*calculateDSum()*' where the sum is calculated and later that method is called in the switch case. This way there is less burden on the 'Hidden' case and this results in a better overall performance of the class.

**<u>Change 1/2: Create new method 'calculateDSum()'</u>**

The case 'Hidden' has a 'for' loop that calculates the sum. This consumes lot of time. Creating a new method will resolve this issue because the 'for' loop can be eliminated and the result can by passed by call to the object calling this new method.

Hence we created 'calculateDSum()' method which does the work of calculating the sum. The result is then passed when the method is called via an object declared in 'Hidden' case.

<u>Diff:</u>

```
public final void train(final double pdExpected, final double pdAlpha, final double pdBeta)
	{
		switch(this.iType)
		{
case HIDDEN:
			{

				double dSum = 0.0;

				for(int i = 0; i < this.oOutputs.size(); i++)
					{
					dSum +=

	((Neuron)this.oOutputs.get(i)).dDelta *

	((Neuron)this.oOutputs.get(i)).getWeight(this);
					}

				this.dDelta = this.dResult * (1.0 - this.dResult) * dSum;
				break;
			}
}
```

**<u>Change 2/2: Move the code for calculating the sum from '*Hidden*' case to method 'calculateDSum()'</u>**

A new method '*calculateDSum()*' is created which removes the burden of the switch case 'Hidden' in calculating the sum. This method is in turn called in the 'Hidden' case of *void train()* using an object.

<u>Diff showing line changes:</u>

```
case HIDDEN:
			{
-				double dSum = 0.0;
-
-				for(int i = 0; i < this.oOutputs.size(); i++)
-				{
-					dSum +=
-
	((Neuron)this.oOutputs.get(i)).dDelta *
-
	((Neuron)this.oOutputs.get(i)).getWeight(this);
-				}
+				double dSum = calculateDSum();
				this.dDelta = this.dResult * (1.0 - this.dResult) * dSum;
				break;
@@ -271,6 +264,18 @@
		}
	}
+	private double calculateDSum() {
+		double dSum = 0.0;
+
+		for(int i = 0; i < this.oOutputs.size(); i++)
+		{
+			dSum +=
+
	((Neuron)this.oOutputs.get(i)).dDelta *
+
	((Neuron)this.oOutputs.get(i)).getWeight(this);
+		}
+		return dSum;
+	}
+
```

2. NeuralNetwork.java

**<u>Change 0/2: Refactor long method 'createlinks()' of Class NeuralNetwork.java to smaller method</u>**

Class NeuralNetwork.java has a long method '*createlinks()*'. This resulted in low cohesion not only in method-level but also in class' structure-level due to the complexity in the code.

Hence there was a necessity to break down this long method into four more methods which has its individual task namely:

- create input link
- create output link
- process neuron
- process layer

The above tasks gave way to the four new methods where each method will take control of the individual work. This is done by taking out the rationale code from ‘createlinks()’ method and delegating it in four methods, namely:

- processLayer()
- processNeuron()
- createInputLink()
- createOutputLink()

This results in improving the cohesion level in both method and class-structure levels.

**Change 1/2: Method ‘createlinks()’ is split into 5 methods**

The original method ‘createlinks()’ is long and complex. There is less cohesion among the inter-communicating methods. In order to increase the cohesion, the long method had to be split further into four more methods, namely:

- processLayer()
- processNeuron()
- createInputLink()
- createOutputLink()

Diff:

```
private final void createLinks(Node poNode)
        throws ClassificationException
        {
                int iType = poNode.getNodeType();
                String strName;
                if(iType == Node.ELEMENT_NODE)
                {
                        strName = poNode.getNodeName();
                        NamedNodeMap oAtts = poNode.getAttributes();
                        if(strName.equals("layer"))
                        {
                                .
                                .
                                .
                        }
                        else if(strName.equals("neuron"))
                        {
                                .
                                .
                                .
                        }
                        else if(strName.equals("input"))
                        {
                                .
                                .
                                .
                        }
                        else if(strName.equals("output"))
                        {
                                .
                                .
                                .
                        }
                // Recurse for children if any
                for
                (
                        Node oChild = poNode.getFirstChild();
                        oChild != null;
                        oChild = oChild.getNextSibling()
                )
                {
                        createLinks(oChild);
                }
   }
```

**Change 2/2: Create new *methods createOutputLink(NamedNodeMap oAtts), createInputLink(NamedNodeMap oAtts), processNeuron(NamedNodeMap oAtts), and processLayer(NamedNodeMap oAtts)***

In order to decrease the complexity of code in long method ‘*createlinks()*’, we have broken down this method into four more methods, namely:

- processLayer()
- processNeuron()

- createInputLink()
- createOutputLink()

Each method, as the name suggests, performs its specific task. By doing so, the complexity in the method '*createlinks()*' and in turn improves cohesion among the methods and within the class.

```
                              if(strName.equals("layer"))
                                          {
-
          for(int i = 0; i < oAtts.getLength(); i++)
-
          {
-
          Node oAttribute = oAtts.item(i);
-
-
          String strAttName = oAttribute.getNodeName();
-
          String strAttValue = oAttribute.getNodeValue();
+
          processLayer(oAtts);
+                                         }


-
          if(strAttName.equals("type"))
-
          {
-
                    if(strAttValue.equals("input"))
-
                    {
-
                              this.oCurrentLayer          =
this.oInputs;
-
                              this.iCurrenLayer = 0;
-
                    }
-
                    else if(strAttValue.equals("output"))
-
                    {
-
                              this.oCurrentLayer          =
this.oOutputs;
-
                              this.iCurrenLayer           =
this.oLayers.size() - 1;
-
                    }
-
                    else
-
                    {
-
                              this.iCurrenLayer           =
++this.iCurrLayerBuf;
-
                              this.oCurrentLayer          =
(Layer)this.oLayers.get(this.iCurrenLayer);
-
                    }
+                                         else
if(strName.equals("neuron"))
+                                         {
+
          processNeuron(oAtts);
+                                         }


-
                    //Debug.debug("Moving to layer " +
currLayer + " [currLayerBuf is " + currLayerBuf + "]");
-
          }
-
          }
+                                         else
if(strName.equals("input"))
+                                         {
+
          createInputLink(oAtts);
                                          }


-                                         else
if(strName.equals("neuron"))
+                                         else
if(strName.equals("output"))
                                          {
-
          String strIndex = new String();
+
          createOutputLink(oAtts);
+                                         }
```

```
+ }

- for(int i = 0; i < oAtts.getLength(); i++)
- {
- Node oAttribute = oAtts.item(i);
+ // Recurse for children if any
+ for
+ (
+ Node oChild = poNode.getFirstChild();
+ oChild != null;
+ oChild = oChild.getNextSibling()
+ )
+ {
+ createLinks(oChild);
+ }
+ }
```

GIPSY

1. TranslationParser.java

**<u>Change 0/2: Refactor TranslatoeParser.java class by introducing a medial method</u>**

The ‘if’ condition increases the complexity of the entire class making it necessary to refactor as it also affects the cohesion (low cohesion). Hence we introduced a medial condition method which uses switch case to decrease the complexity.

**<u>Change 1/2: Splitting *‘HashtableParse()’* method into two smaller methods</u>**

In order to reduce the complexity of code and improve cohesion in class TranslatorParser, a new *‘medialConditions()’* method is created which replaces the complex ‘if’ condition and instead uses a simple switch case.

```
public Hashtable Parse() throws IOException
{
.
.
.
if (strTemp.charAt(1) == '1')
{
DoCond();
}
else if (strTemp.charAt(1) == '2')
{
DoThen();
}
else if (strTemp.charAt(1) == '3')
{
DoElse();
}
else if (strTemp.charAt(1) == '4')
{
DoAt();
}
else if (strTemp.charAt(1) == '5')
{
DoEqual();
}
else if (strTemp.charAt(1) == '6')
{
DoBrace();
}
else if (strTemp.charAt(1) == '7')
{
DoNotBrace();
}
}
```

**Change2/2: Create new method medialConditions()**

A new method called as ‘medialConditions()’ method is created, which replaces the complex ‘if’ condition of ‘HarseParse()’ method into a simple switch case that reduces the complexity of code, thereby improving the structure of the class and improving the cohesion.

```
+       public void medialConditions(char choice) {
+               if (choice == '1')
+               {
+                       DoCond();
+               }
+               else if (choice == '2')
+               {
+                       DoThen();
+               }
+               else if (choice == '3')
+               {
+                       DoElse();
+               }
+               else if (choice == '4')
+               {
+                       DoAt();
+               }
+               else if (choice == '5')
+               {
+                       DoEqual();
+               }
+               else if (choice == '6')
+               {
+                       DoBrace();
+               }
+               else if (choice == '7')
+               {
+                       DoNotBrace();
+               }
+       }
+
```

2. Configuration.java

**Change 0/1: Simplify conditional expression using Refactoring for equals method**

The if else condition in equals method of configuration class is simplified using refactoring. The method call implemented is very complex. We tried to make it efficient by creating some objects. Thus, the method call becomes simpler.

**Change 1/1: Make method call simple**

There are two objects namely setting and isValidConfig which are used to store the result from different method calls. The diff for this change is shown below. It shows all the changes in the code.

```
public boolean equals(Object pObject)
        {
+               boolean isValidConfig = false;
+
                if(pObject instanceof Configuration)
                {
-                       return this.oConfigurationSettings.equals(((Configuration)pObject).oConfigurationSettings);
+                       Object setting = ((Configuration)pObject).oConfigurationSettings;
+                       isValidConfig = this.oConfigurationSettings.equals(setting);
+
                }
-               else
-               {
-                       return false;
-               }
-       }
+
+               return isValidConfig;
+       }
+
```

### Test Cases

We have implemented two test cases which are shown below. These unit test cases are used to show that external behaviour of the system is not changed. Thus, refactoring is implemented successfully.

i. Test Case 1: GlobalInstanceTest

```
package gipsy.tests;
import static org.junit.Assert.*;
import junit.framework.Assert;
import
gipsy.RIPE.editors.RunTimeGraphEditor.core.GIPSYInstance;
import
gipsy.RIPE.editors.RunTimeGraphEditor.core.GIPSYPhysicalNode;
import
gipsy.RIPE.editors.RunTimeGraphEditor.core.GIPSYTier;
import
gipsy.RIPE.editors.RunTimeGraphEditor.core.GlobalInstance;
import
gipsy.RIPE.editors.RunTimeGraphEditor.core.NodeConnection;
import org.junit.Test;
import static org.junit.Assert.*;
import org.junit.Test;
// Junit class to test Global Instance
public class GlobalInstanceTest {
	// Testing isItemExists method
	@Test
	public void isItemExistsTest()
	{
		GIPSYPhysicalNode GipsyPhysicalNode = new GIPSYPhysicalNode();
	GlobalInstance.getInstance().addGIPSYPhysicalNode(GipsyPhysicalNode);
		Assert.assertEquals(true, GlobalInstance.getInstance().isItemExists(GipsyPhysicalNode));
		GIPSYInstance GipsyInstance = new GIPSYInstance();
	GlobalInstance.getInstance().addGIPSYInstance(GipsyInstance);
	Assert.assertEquals(true,GlobalInstance.getInstance().isItemExists(GipsyInstance));
		GIPSYTier GipsyTier = new GIPSYTier();
	GlobalInstance.getInstance().addGIPSYTier(GipsyTier);
	Assert.assertEquals(true,GlobalInstance.getInstance().isItemExists(GipsyTier));
		NodeConnection nodeConnection = new NodeConnection();
	GlobalInstance.getInstance().addNodeConnection(nodeConnection);
	Assert.assertEquals(true,GlobalInstance.getInstance().isItemExists(nodeConnection));
	}
}
package gipsy.tests;
```

ii. Test Case 2: ConfigurationTest

```
import static org.junit.Assert.*;
import gipsy.Configuration;
import org.junit.Assert;
import org.junit.Test;

//Junit class to test Configuration
public class ConfigurationTest {
	// Testing equals method
	@Test
	public void equalsTest()
	{
		// Create First object of configuration class
		Configuration gipsyConfig = new Configuration();
		// Create Second object of configuration class
		Configuration gipsyConfig2 = new Configuration();
		// Checks weather objects are equal

	Assert.assertEquals(gipsyConfig,gipsyConfig2);
	}
		}
```

## VII. Conclusion

To conclude, having a good software design and architecture will directly affect the quality of the software system (such functionality, usability, reliability, efficiency, reusability, extendibility, and maintainability). In our study we have first shown a general overview about the two systems under investigation, DMARF and GIPSY. After we start identifying the different factors related to quality attributes (Analyzability, Changeability, Stability, and Testability) and their measures using different tools. We have used the LogiScope and McCabe tools to generate the list of classes having code smells and poor design quality. After we have chosen some classes having poor design attributes and code smells and we have implemented some refactoring on these classes to improve their measures and quality. Following some design patterns and known types of refactoring we have implemented and modified the code of these classes mentioned previously. Of these types we mention: pull up method, removing clone (duplicate) code, simplifying conditional expressions, and make method calls simpler. Other than that we have identified and described some known design patterns that are used in the design of the system (such as Factory patterns).

As a part of future work on the two systems working on refactoring the other list of classes with design issue will definitely improve the software quality. Another extension to the existing systems would be a fused system where both systems are merged in such a way that we run DMARF on the GIPSY's runtime (mainly the GEE subsystem) for a distributed computing at the runtime.

## VIII. Glossary

| Term | Definition |
|---|---|
| Classifier | That indicates the semantic class to which a noun belongs |
| Data graph | It is graphical version of design. |
| Demand Driven Model | Caused or determined by demand from consumers or clients. |
| Demand Generator | The things that drives someone to make certain purchases. |
| Distributed Multi-Tier | The same processing is replicated over several nodes. In the other case, each tier has a distinct responsibility and the processing running on each tier differ. |
| Evaluator | Sends the data to the compiler which will finally compute the final result. |
| Feature Extractor | Transforming the input data into the set of features |
| GIPSY Node | registered process that hosts one or more GIPSY tier instances belonging to different GIPSY instance |
| Loaders | A loader is the part of an operating system that is responsible for loading programs. |
| Lucid intentional programming | Lucid uses a demand-driven model for data computation |
| Multi-tier architecture | In software engineering, multi-tier architecture (often referred to as *n*-tier architecture) is a client–server architecture in which presentation, application processing, and data management functions are physically separated. |
| Neural Network | A computer system modeled on the human brain and nervous system. |
| Node Controller | It controls the nodes. |
| Pattern recognition | Pattern recognition refers to the process of recognizing a set of stimuli arranged in a certain pattern that is characteristic of that set of stimuli. Pattern recognition does not occur instantly, although it does happen automatically and spontaneously. |
| Pipeline | a linear sequence of specialized modules used for pipelining |
| Preprocessing | Subject (data) to preliminary processing. |
| Registered node | Every node type registered with the repository has a unique name |
| Sample Loader | loader loads the data from application |
| Self-healing | The system must effectively recover when a fault occurs, identify the fault, and, when possible, repair it. |
| Self-optimization | An AC system can measure its current performance against the known optimum and has defined policies for attempting improvements. |
| Self-protection | The system must defend itself from accidental or malicious external attacks, |
| Speaker Identification | It identifies the speaker's voice in application. |
| Storage Manager | Provide data management for distributed computing. |
| Web servlet | define servlets as a part of a Web application in several entries |

*Table 13: Glossary*